\documentclass[%
 reprint,
superscriptaddress,
 amsmath,amssymb,
 aps,
prx,
]{revtex4-2}
\usepackage{comment}
\usepackage{lipsum}
\usepackage{booktabs}
\usepackage{graphicx}
\usepackage{dcolumn}
\usepackage{array}
\usepackage{bm}
\usepackage{hyperref}
\usepackage[mathlines]{lineno}
\usepackage{xcolor}
\usepackage[normalem]{ulem}
\usepackage{braket}
\usepackage{multibib} 

\newcites{app}{Appendix References} 

\definecolor{doubt}{RGB}{0, 102, 204} 
\definecolor{author2}{RGB}{204, 0, 0}   
\definecolor{review}{RGB}{34, 139, 34}  

\newcommand{\Ca}{$^{40}$Ca$^+$\,}

\newcommand{\Ybf}{$^{171}$Yb$^+$\,}
\newcommand{\Ybc}{$^{172}$Yb$^+$\,}
\newcommand{\Yb}{Yb$^+$\,}
\newcommand{\Ba}{Ba$^+$\,}

\newcommand{\Rice}{Department of Physics and Astronomy and Smalley-Curl Institute, Rice University, Houston, TX 77005, USA}
\newcommand{\Duke}{Duke Quantum Center and Department of Physics, Duke University, Durham, NC 27701, USA}
\newcommand{\UMD}{Joint Quantum Institute and Department of Physics, University of Maryland, College Park, MD 20742, USA}
\newcommand{\APP}{Applied Physics Graduate Program, Smalley-Curl Institute, Rice University, Houston, TX 77005, USA}
\newcommand{\Translume}{Translume Inc., Ann Arbor, MI 48108, USA}
\newcommand{\TAMOS}{TAMOS Inc., Houston, TX 77004, USA }
\newcommand{\QLAB}{National Quantum Laboratory (QLab), University of Maryland, College Park, MD 20742, USA}

\begin{document}

\preprint{APS/123-QED}
\raggedbottom

\title{Monolithic Segmented 3D Ion Trap for Quantum Technology Applications}

\author{Abhishek Menon}
\email{am219@rice.edu}
\affiliation{\Rice}
\author{Michael Straus}
\affiliation{\Duke}
\affiliation{\UMD}
\author{George Tomaras}
\affiliation{\Rice}
\affiliation{\APP}
\author{Liam Jeanette}
\affiliation{\Duke}
\affiliation{\UMD}
\author{April X. Sheffield}
\affiliation{\Rice}
\author{Devon Valdez}
\affiliation{\Duke}
\author{Yuanheng Xie}
\affiliation{\Duke}
\affiliation{\UMD}
\author{Visal So}
\affiliation{\Rice}
\author{De Luo}
\affiliation{\Duke}
\author{Midhuna Duraisamy Suganthi}
\affiliation{\Rice}
\affiliation{\APP}

\author{Mark Dugan}
\affiliation{\Translume}
\author{Philippe Bado} 
\affiliation{\Translume}
\author{Norbert M. Linke}
\email{linke@umd.edu}
\affiliation{\Duke}
\affiliation{\UMD}
\affiliation{\QLAB}
\author{Guido Pagano}
\email{pagano@rice.edu}
\affiliation{\Rice}
\affiliation{\TAMOS}
\author{Roman Zhuravel}
\email{zhuravel@tamosolutions.com}
\affiliation{\Rice}
\affiliation{\TAMOS}

\begin{abstract}
Monolithic three-dimensional (3D) Paul traps combine the high-precision microfabrication of two-dimensional (2D) chip traps with the deep trapping potentials and low heating rates characteristic of macroscopic 3D Paul traps, which are typically machined by traditional means and mechanically assembled. However, achieving low motional heating rates and optical access with a high numerical aperture (NA) while maintaining the high radio-frequency (RF) voltages required for trapping heavy ionic species, such as $\text{Yb}^{+}$ and $\text{Ba}^{+}$, remains a significant technical challenge. In this work, we present a fused-silica, monolithic segmented 3D Paul trap with an ion-electrode distance of $250~\mu\text{m}$, and stable operation at high RF voltages. We benchmark the trap's performance using $\text{Yb}^{+}$ ions, demonstrating axially homogeneous trapping potentials spanning over 200 $\mu\rm m$ about the trap's axial center, high multi-directional optical access (up to $0.7$ NA),
and radial motional heating as low as $\dot{\bar n}=1.1\pm0.1\ \mathrm{quanta/s}$ at radial trap frequencies about $3\ \mathrm{MHz}$ near room temperature. Furthermore, we observe a motional Ramsey coherence time, $T_2$, of about $95~\text{ms}$ for the radial center-of-mass mode. We demonstrate the generation of a two-qubit Bell state with a parity contrast
of $99.3^{+0.7}_{-1.5}$ \% with state preparation and measurement correction. These results establish fused-silica monolithic 3D Paul traps as a scalable, modular platform for quantum simulation, computation, metrology, and networking with heavy ionic species. 
\end{abstract}

\maketitle

\section{\label{sec_intro} Introduction}

Trapped ions have emerged as one of the leading technologies for quantum computing \cite{Bruzewicz2019}, simulation \cite{monroe2021programmable}, networking \cite{Duan2010}, and metrology \cite{Ludlow2014} over the past few decades. The design and performance of their backbone--the ion trap itself--continue to advance and are crucial for each application. Radio-frequency (RF) Paul traps \cite{Paul1958Forsch, paul1990electromagnetic} and static magnetic-field based Penning traps \cite{Brown1986Geonium} form the two parent branches of ion traps, using confinement from a ponderomotive RF force or a Lorentz force, respectively, in addition to static electric fields. 
Historically, ion traps have been traditionally machined and mechanically assembled, making them prone to misalignment, lacking reproducibility and scalability, and offering limited geometries. However, advances in microfabrication techniques \cite{romano2025etching, TranslumeWebsite,basu2025advancements, xie2019laser} have broadened the capabilities of both Paul traps \cite{maunz2016HOA,sterk2024multi, moses2023race} and Penning traps \cite{Jain2024} by facilitating Quantum Charged-Coupled Device (QCCD) architectures in 2D chip traps \cite{pino2021}, and integrating optics and detectors into the trap \cite{Niffenegger2020, Mehta2020, Ivory2021, Kwon2024, Mordini2025}. More recently, these techniques have raised interest in microfabricated 3D Paul traps \cite{wilpers2012monolithic, Xu20253D, auchter2022industrially, jordan2025scalable, Lovera2024}.

Monolithic 3D Paul traps reap the benefits of both traditional macroscopic 3D Paul traps and modern monolithic 2D chip traps. Their microfabricated 3D structures can host deep, symmetric trapping potentials, with multi-directional optical access, trap depth in the eV range, and low heating rates owing to large ion-electrode distances \cite{turchette2000heating}. 
The fabrication process enables manufacturing repeatability with micron-level precision 
\cite{maunz2016HOA, dugan2015microfabrication}, thereby facilitating segmented electrodes for zone-based quantum operations \cite{Lovera2024} and providing stable, reconfigurable, trapping potentials.
These features make monolithic 3D Paul traps a scalable and modular platform for quantum technology applications \cite{Xu20253D}. Furthermore, monolithic 3D traps have lately become a key ingredient in the manipulation of large 2D ion crystals for quantum simulation \cite{Wang2020coherently, Kiesenhofer2023, Qiao2024tunable, Guo2024site-resolved, Guo2025Quantum}, and progress has been made towards their integration with cavities for quantum networking \cite{teh2024ion,kassa2025integrate}.

While these desirable features of monolithic 3D Paul traps have been verified individually, no platform has demonstrated all of them simultaneously. Crucially, there have been no demonstrations of monolithic 3D traps capable of manipulating heavy ionic species, such as Ba$^+$ and Yb$^+$, which are workhorse atomic systems in trapped-ion quantum information science, while retaining all the features listed above. These ionic species require shorter ion-electrode distances, $d$, and/or higher trap RF voltages to achieve $\sim$ 3-4 MHz frequencies compared to their lighter counterparts, such as Be$^+$ or Ca$^+$. This requirement leads to significant technical and fabrication challenges \cite{brown2021materials}: greater RF isolation is required to prevent dielectric breakdown, and the chosen substrate must support high surface quality, reproducible microfabrication, and good thermal and electrical properties. Additionally, reducing the ion-electrode distance leads to an unfavorable scaling of heating rates $(\sim1/d^4)$ \cite{Brownnutt2015}, which, in turn, degrades motional mode coherence   \cite{turchette2000decoherence,talukdar2016implications}, crucial for quantum information science \cite{Fluhmann2019encoding, matsos2025universal}, metrology \cite{sagesser20243,biercuk2010ultrasensitive}, and precision spectroscopy \cite{Ludlow2014}.

In this work, we report the first segmented monolithic 3D trap capable of manipulating heavy ionic species with a low heating rate ($\sim 1$ quanta/s) at an ion-electrode distance of $d=250\,\mu$m, long motional Ramsey coherence times ($T_2\sim 95$ ms),  and high multi-directional optical access (up to 0.7 numerical aperture, NA).
The trap is fabricated by Translume Inc. using their \emph{femtoEtch}$^{\emph{\text{TM}}}$ process of Selective-Laser Etching (SLE) on a 2 mm thick substrate of fused silica \cite{bellouard2003monolithic, dugan2015microfabrication}.
We benchmark the trap's performance with \Yb ions, demonstrating robust, stable operation at high RF voltages ($\gtrsim 450$ V$_{\rm{pk}}$) with axially homogeneous radial trapping potentials, $\dot{\bar{n}}<10$ q/s for both radial modes, high photon collection efficiency, and low excess micromotion.
We also report the effects of \emph{ex situ} plasma cleaning on an ion trap, which may contribute to the observed 100-fold reduction in heating rates compared to our previous non-surface-treated trap designs. 
Crucially, this places our room-temperature trap performance on par with typical cryogenic ion traps \cite{Bruzewicz2015}.

The paper is structured as follows:  In Sec. \ref{sec_designandassembly}, we motivate the general design and material choices of the monolithic trap assembly, including its thermal and electrical properties. 
In Sec. \ref{sec_trapfreqMM}, we report the characterization of the trapping potential and the axial micromotion using Yb$^+$ ions as the probe. In Sec. \ref{sec_heatingrate_trends}, we present measurements of the heating rate of both radial motional modes. In Sec. \ref{sec_coherentmanip}, we show coherent manipulation of the trapped-ion qubit and its motion. 
Finally, in Sec. \ref{sec_discussion}, we discuss how the features of this monolithic ion trap can enable many quantum technology applications.

\section{\label{sec_designandassembly} Design and Assembly}

\begin{figure*}[t!]
    \centering
    \includegraphics[width=\textwidth]{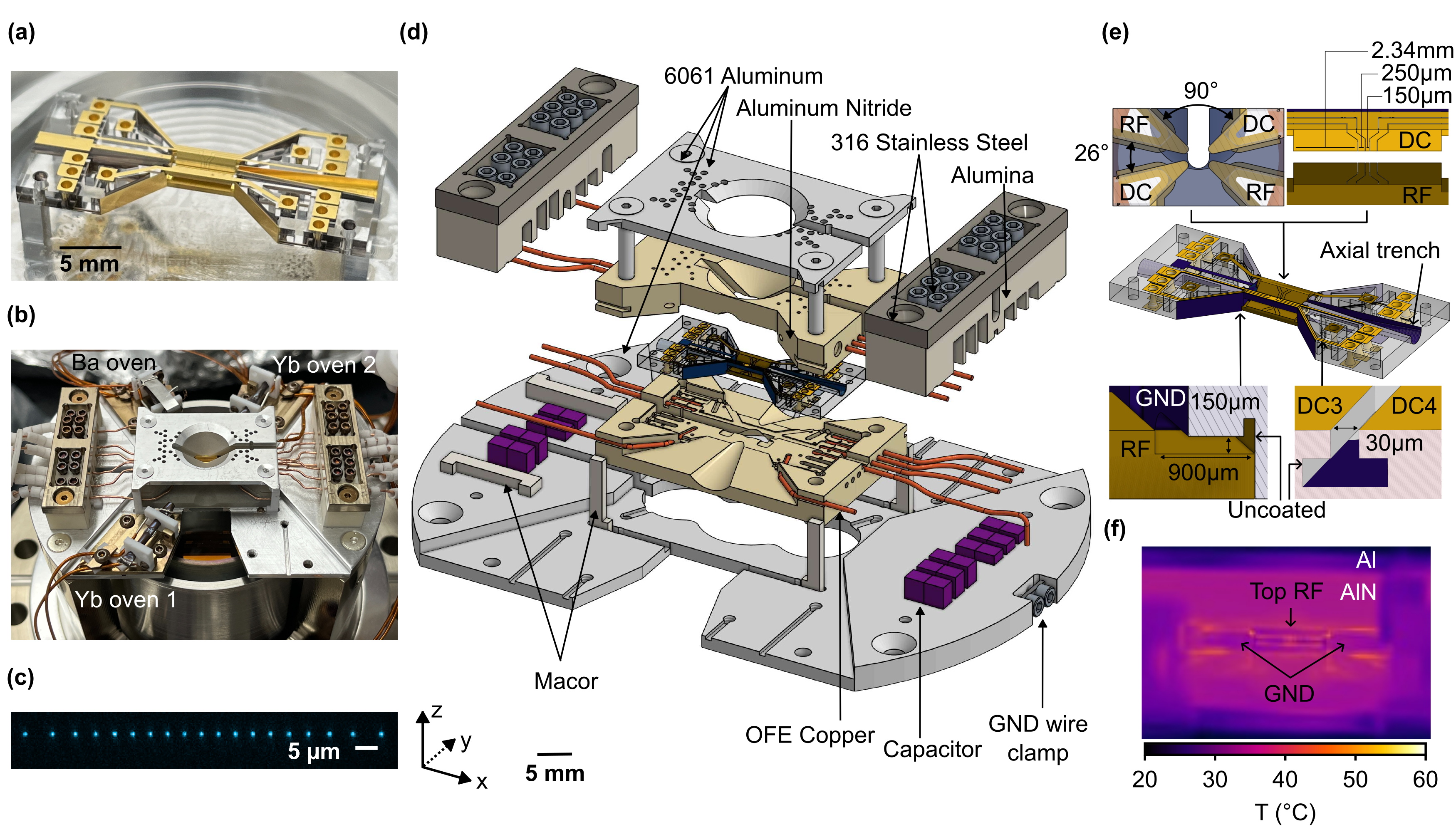}
    \caption{ 
    {\bf (a)} 
    GEN3 segmented, monolithic 3D blade trap made of fused silica, with $d=250\,\mu$m.
    {\bf (b)} The monolithic trap assembly and the oven assembly, for Yb$^+$ and Ba$^+$ ions, before being inserted into the vacuum chamber. 
    {\bf (c)} A Doppler-cooled linear crystal of 19, $^{171}\rm{Yb}^{+}$ ions at $\omega/2\pi\sim3$ MHz radial confinement, with an inter-ion spacing of $4.2(3)\,\mu$m for the central 13 ions, for which the uniformity was optimized with 7\% variability.
    {\bf (d)} An exploded view of the trap stack detailing the metal and ceramic layers for heat dissipation, and the routing of 22 AWG copper wires from the trap's gold pads (through fuzz-buttons, not shown) to the in-vacuum filtering capacitors for the DC electrodes. The ground wire connects the trap's ground pad to the aluminum base plate.
    {\bf (e)} Insets of critical features of the trap, which include the high multi-directional optical access from the blade geometry and axial trenches, segmented electrodes, and trench geometries for DC-DC and RF-GND isolation.
    {\bf (f)} Thermal image of the trap stack under high-power RF test in high vacuum conditions. The color-coded temperature map is measured at $V_{\rm{RF}}=680 \ \rm{V_{pk}}$, $\Omega_{\rm{RF}}/2\pi=23.26\ \rm{MHz}$. The thermal IR camera temperature reading is calibrated to an in-vacuum thermocouple, in contact with the top AlN ceramic of the trap stack.
    }
    \label{fig_trapgen}
\end{figure*}

The general design of our monolithic 3D ion trap is inspired by the macroscopic blade trap design used in Refs. \cite{debnath2016demonstration, Pagano2018, so2024trappedion}, miniaturized and laser written into a $30.5 \times 13 \times 2 \,{\rm mm}^3$ substrate of fused silica ($a$-SiO$_2$). As shown in Fig.~\ref{fig_trapgen}a, the gold-coated blades of the trap constitute the neck of a bowtie with dimensions $\rm{6 \times 3.5 \times2\  mm^3}$. 
The trap is sputter-coated with 2 $\mu$m of gold on top of a 100 nm adhesion layer of titanium. The blade geometry features an ion-electrode distance of $d=250\,\mu$m, unless otherwise noted. The blades are inclined to 26$^\circ$ with respect to the $y$ axis to provide a bi-directional solid angle of 0.23 NA along the $y$ axis and 0.7 NA along the $z$ axis from the trap center, enabling multi-directional individual-ion addressing and high photon collection efficiency. The blade tips have a 95 $\mu$m radius of curvature to avoid sharp corners, which would generate large local electric fields. The tapered axial trench through the trap axis (see Fig.~\ref{fig_trapgen}b,d,e) provides a path for service beams, which, if needed, can be angled vertically up to $\sim19^{\circ}$, to improve cooling and provide access to a micromotion compensation beam with projection along the vertical $z$ direction. The two opposing DC blades are each segmented into five independent electrodes, which we refer to in sequence as: endcap, midcap, centercap, midcap, and endcap. The two opposing RF blades are each a single electrode but feature the same segmentation geometry as the DC blades at the blade tips to provide symmetry and reduce axial micromotion. The length of the centercaps ($150\ \mu\rm{m}$) and midcaps ($250\ \mu\rm{m}$) were specifically chosen to minimize the axial inhomogeneity in the radial trap frequency along the trap axis $x$ (see Fig.~\ref{fig_trapgen}e, ~\ref{fig_trapfreqtrends}a) and to minimize the voltages required for anharmonic axial potentials to trap ion chains with equally spaced ions \cite{Johanning2016, Pagano2018} (see Fig.~\ref{fig_trapgen}c). This enables operations with well-behaved collective modes \cite{home2011normal} and low crosstalk for individual-ion addressing, cooling, and detection \cite{Lin_2009, egan2021scaling}.

The electrode isolation is achieved by laser-writing three-dimensional trenches of different dimensions in the $a$-SiO$_2$ substrate \cite{Kiesenhofer2023}. Each electrode is routed to a gold pad and connected to the RF and DC sources via fuzz-button interconnects and bare copper wires inside the vacuum chamber. Using bare copper wires and ceramic connectors avoids any soldering or wire-bonding and minimizes background vacuum pressure. Each DC electrode is filtered by a pair of in-vacuum $\sim$ 1.25 nF capacitors, which are fit into vented pockets in the aluminum base-plate of the trap stack (see Fig.~\ref{fig_trapgen}b,d). Each of them, including the DC bias on the RF electrodes, is further filtered with a low-pass $\pi$-filter (LCR ladder network), with a cutoff $<1$ kHz, outside the vacuum system.

To ensure a stable trapping potential at high RF voltages, the ideal trap substrate would feature a precisely fabricated structure with a low coefficient of thermal expansion and a high Young's modulus to minimize thermal and mechanical deformations. A high thermal conductivity also facilitates rapid heat-redistribution and heat-transfer from the trap to a colder reservoir. Further, materials with a low loss tangent and a low dielectric constant, which minimizes the trap capacitance, help to reduce the power dissipated on the trap.

High ionization thresholds are also desirable to minimize the possibility of charging of the unmetallized parts of the trap exposed to UV laser light. 
A high dielectric strength ensures high voltage breakdown thresholds between RF and DC (or Ground) electrodes across the substrate. Moreover, a high vacuum surface flashover (VSF) strength is desirable since it can be an order of magnitude smaller than the bulk dielectric strength via the `triple-junction effect' in vacuum 
\cite{zhang2025global, kumar2025vacuum, pillai1985surface, miller1989surface} (see Appendix \ref{app_thermtestdata}).

Diamond and sapphire have most of the favorable characteristics (see Table~\ref{tab:trap_materials}
in Appendix \ref{app_thermtestdata}), but they are challenging to microfabricate into three-dimensional, monolithic geometries due to their chemical and crystalline properties \cite{tan2022femtosecond, ali2021femtosecond, cole2023high}. We choose fused silica as a good alternative for our monolithic trap because it is a well-established substrate in Selective-Laser Etching \cite{dugan2015microfabrication, bellouard2003monolithic}. Fused silica features one of the lowest tangent-loss coefficients ($\sim10^{-4}$), dielectric constants, and thermal coefficients of expansion, along with a high Young's modulus (similar to that of aluminum) and a high ionization threshold. Hence, it can provide a stable, dimensionally precise, and rigid structure under high RF power. To compensate for fused silica's low thermal conductivity, the trap is enclosed in two layers of aluminum nitride (AlN) with $\sim$150 W/m$\cdot$K thermal conductivity, which are in turn connected to two aluminum layers that act as heatsinks (see Fig.~\ref{fig_trapgen} b,d) while providing additional electrical shielding. 

We use narrow, inverted T-shaped trenches to isolate the DC electrodes from one another and wider, U-shaped trenches to isolate the RF electrodes from the DC (or Ground) electrodes (Fig.~\ref{fig_trapgen}e). 
Importantly, the trench shapes are optimized to reduce capacitance between the RF and DC (or Ground) electrodes and increase the VSF breakdown voltage \cite{zhang2025global}.
This is crucial to reduce the presence of hot-spots and minimize the power dissipated on the trap \cite{Stick2006, dietl2025test, Siverns2012OnResonators}:
\begin{equation}
    \centering
    \label{eqn:RF dissip}
    P_{\rm{dissip}}=\frac{1}{2}R_s\ C_t^2\ \Omega_{\rm{RF}}^2 V_{\rm{RF}}^2.
\end{equation}
where $R_s$ is the effective series resistance of the electrodes, $C_t$ is the net capacitance of the trap
, $\Omega_{\rm{RF}}/2\pi$ is the RF drive frequency, and $V_{\rm{RF}}$ is the RF voltage amplitude.

The heat dissipation is mostly attributed to ohmic processes, given its negligible contribution from the tangent loss of fused silica.
To optimize the thermal properties of the trap assembly, we characterize the trap at high RF power under high vacuum conditions and image the trap on one side with a FOTRIC 346A-L25 thermal IR camera through a ZnSe window \cite{Hainzer2023, nordmann2020subKTraptempstabiliz} while also imaging it from the other side with a visible microscope through a fused-silica window. In the first design iteration (referred to as GEN1, see Appendices~\ref{app_projectworkflow},~\ref{app_thermtestdata}), we observed distinct hotspots at the RF-DC isolation trenches that were glowing in the visible spectrum starting from $V_{\rm RF}\sim\ 400$ V$_{\rm{pk}}$. We attribute these hotspots to both the large capacitance of those trenches in earlier designs and to possible VSF between RF and DC electrodes. 
In successive design generations (GEN2, GEN3), we reduced the capacitance of the RF-DC (or Ground) trenches by increasing their shortest isolating dimension from 75 ${\rm\mu m}$ to
100 ${\rm\mu m}$ (150 ${\rm\mu m}$), and changed the geometry from L-shaped to long U-shaped trenches of 700 $\mu$m (900 $\mu\rm{m}$) in depth for the GEN2 (GEN3) trap. This led to the disappearance of visible radiation up to a maximum test voltage of 1000 $\rm{V}_{\rm pk}$ at $\Omega_{\rm{RF}}/2\pi\sim$ 37 MHz, and a significant reduction in local hot-spots and the overall trap stack temperature in successive generations (see Fig.~\ref{fig_trapgen}f) compared to GEN1.

\section{\label{sec_trapfreqMM} Trap potential and Micromotion}

We calculate the static DC potential and RF pseudo-potential using Finite Element Analysis (FEA). The GEN2 trap ($d=250\,\mu$m) and the two variants of the GEN3 trap, GEN3-250$\mu$m ($d=250\,\mu$m) and GEN3-200$ \mu$m ($d=200\,\mu$m), have a radial geometric factor of $\kappa\approx0.83$, which quantifies the deviation from a purely hyperbolic potential \cite{leibfried2003quantum,berkeland1998minimization}, and a trap depth of $\sim$ 1 eV (see Appendix \ref{app_trapsim}).

\begin{figure}[t!]
    \centering
    \includegraphics[width=\columnwidth]{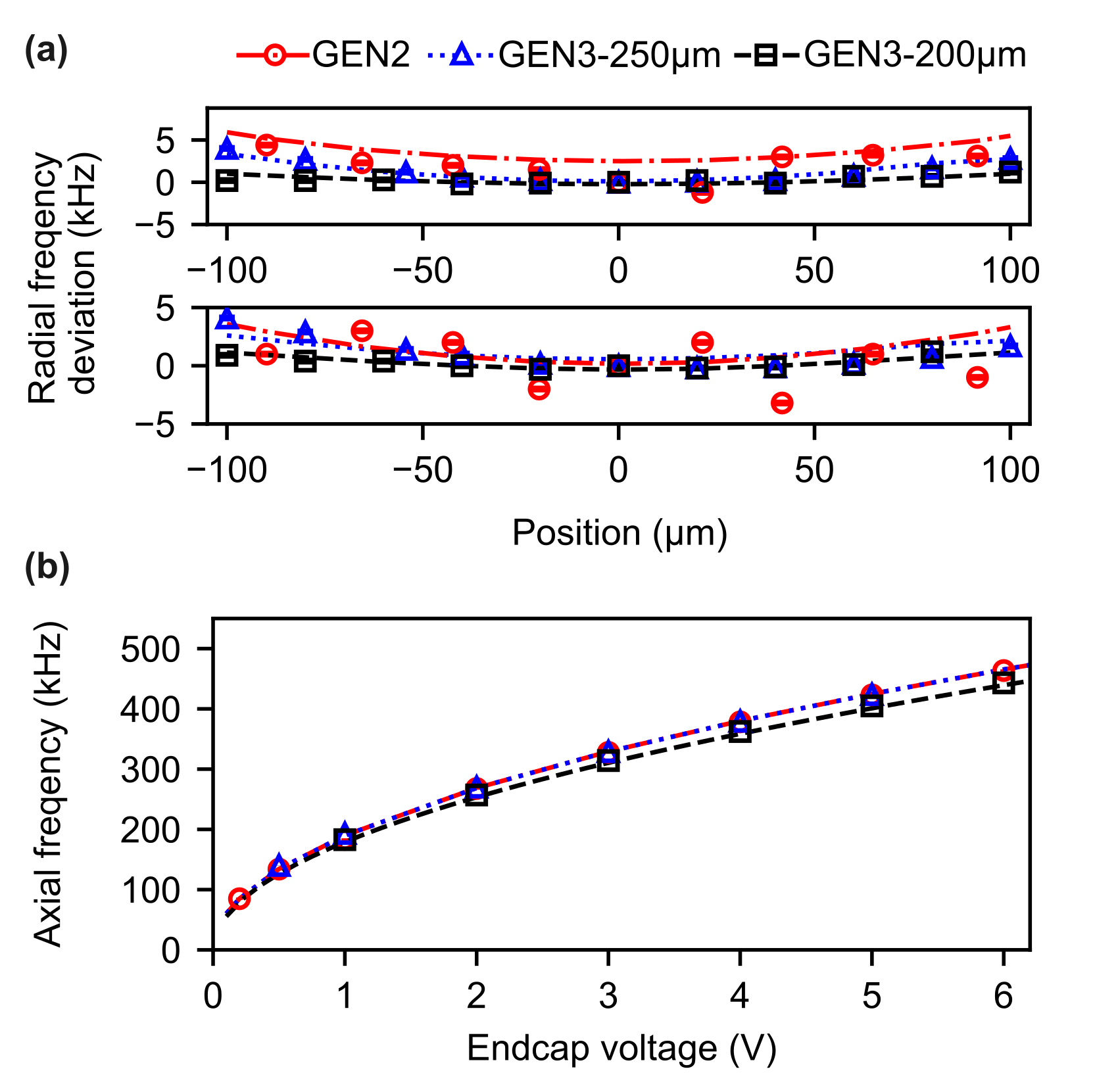}
    \caption{ 
     {\bf (a)} Comparison of radial frequencies measured as a function of position along the $x$ axis for the GEN2 (red), GEN3-$250\mu$m (blue), and GEN3-$200\mu$m (black) traps. Top and bottom panels refer to the HF and LF radial modes, respectively, along with their simulation-based fits (colored lines).
     The radial frequency deviations for the HF (LF) radial mode of the GEN2, GEN3-$250\mu$m, and GEN3-$200\mu$m traps are about an offset of 1.979 MHz (1.932 MHz), 2.899 MHz (2.628 MHz), and 3.038 MHz (2.771 MHz) at the trap center. The RF voltages applied are  $\approx 380\,\rm{V_{pk}},\, 480\,\rm{V_{pk}},\,\text{and}\, 316\,\rm{V_{pk}} $, with `twist' voltages of $-0.3\,\text{V},\,-2\,\text{V},\,\text{and}\, 1.28\,\text{V}$. 
     The shallow concavity in the HF and LF modes is due to the $ x$-axis `push' that shifts the ion, and is well captured by the simulation.
    {\bf (b)} Comparison of the axial frequency measurements of the GEN2 and GEN3 traps vs applied voltage on the endcaps, and their FEA simulations (colored lines).
    }
    \label{fig_trapfreqtrends}
\end{figure}

We measure the radial trap frequencies of a single Yb$^+$ ion at different axial locations by scanning the frequency of the RF signal applied on an external antenna to resonantly drive the ion's motional modes, and observe the ion's spatial excitation on an EMCCD camera (`RF-tickling') \cite{dehmelt1990less}. We shuttle the ion over the span of 200 $\mu$m across the axial center of the GEN2 and GEN3 traps, and compare the measurements with FEA simulations of the trapping potential (Fig.~\ref{fig_trapfreqtrends}a). Using a static quadrupole in addition to the pseudopotential, we can lift the degeneracy between the two radial modes, referred to as the higher center-of-mass radial frequency (HF) and lower center-of-mass radial frequency (LF). This involves biasing the DC (RF) electrodes with a positive (negative) $V_{\rm{twist}}$, referred to as the `twist' voltage. Applying a `twist' also rotates the radial mode principal axes because of the $\sim$2:1 ($y$:$z$) aspect ratio of our blade geometry \cite{Saito2024, johnson2016experiments}.

As shown in Fig.~\ref{fig_trapfreqtrends}a, we observe a root-mean-squared (RMS) fractional deviation of 0.01-0.15\% in the HF (top panel) and LF (bottom panel) radial frequency measurements from the simulation, with measurement uncertainties of $\pm\,$2 kHz, $\pm\,$0.5 kHz, $\pm\,$0.7 kHz for the GEN2, GEN3-$250\mu$m, and GEN3-$200\mu$m trap, respectively.
Since the RF voltage is challenging to measure independently and accurately, we leave it as a fitting parameter for the radial frequency simulation (see Appendices~\ref{app_radialfreqfits},~\ref{app_RFsetup}).

We use the same `RF-tickling' method at the center of the GEN2 and GEN3 traps to measure the axial frequency, as a function of endcap voltage, and compare it with respect to the simulation, without fitting parameters.  
In Fig.~\ref{fig_trapfreqtrends}b, we observe a RMS fractional deviation of 1.3\%, 1.8\%, and 1.4\% for the axial frequencies of the GEN2, GEN3-$250\mu$m, and GEN3-$200\mu$m trap, respectively, from the FEA simulation.
The close agreement of the trap frequencies with the simulation is an indication of the high fabrication precision and accuracy of the SLE microfabrication technique for fused silica, which was tailored to compensate for both laser aberration and finite etching selectivity \cite{dugan2015microfabrication, bellouard2003monolithic, Bellouard2004}.

\begin{figure}[t!]
    \centering
    \includegraphics[width=\columnwidth]{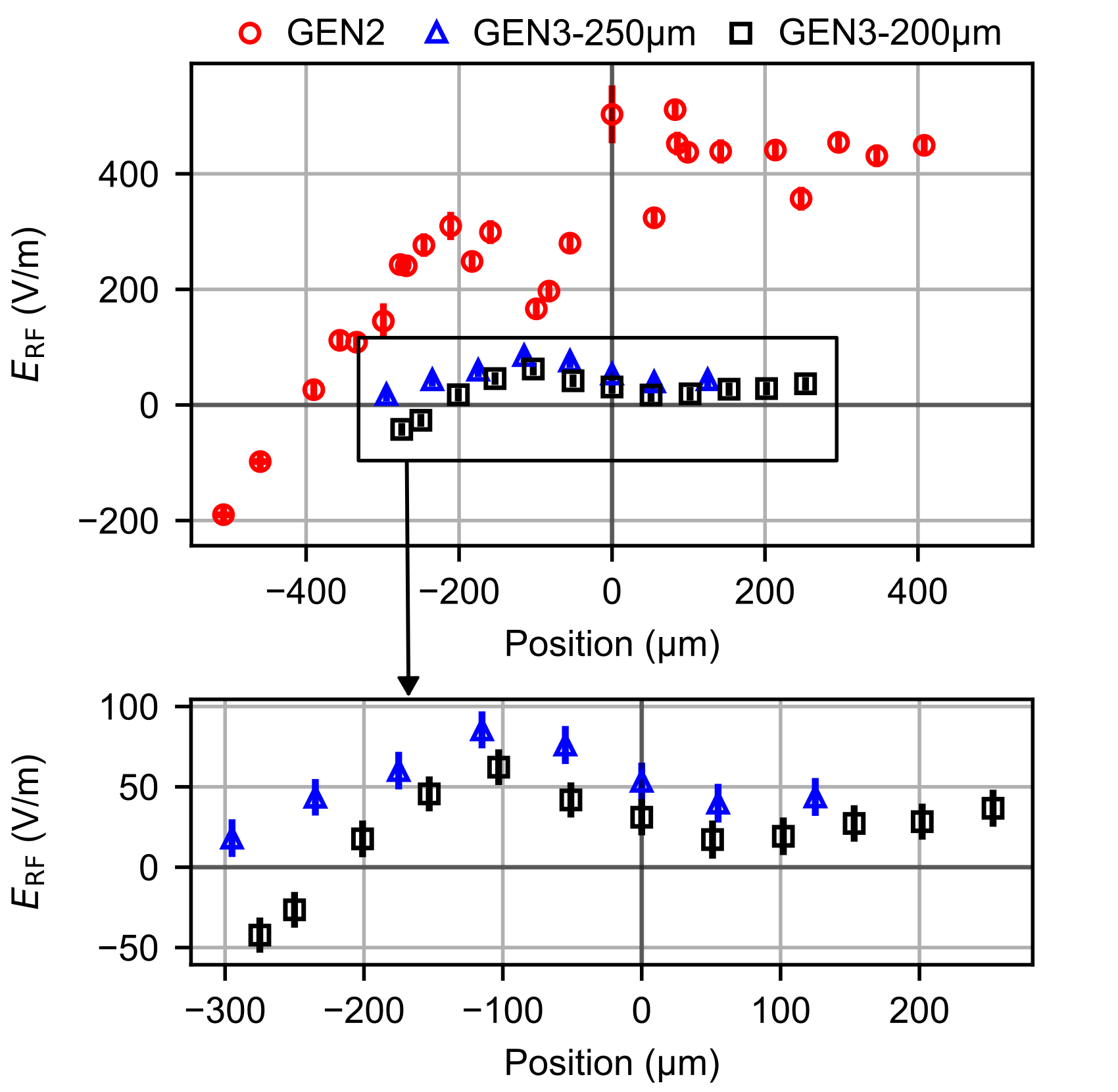}
    \caption{ 
    Residual axial EMM profile of the GEN2 (red), GEN3-$250\mu$m (blue), and GEN3-$200\mu$m (black) traps at $\omega/2\pi\sim3$ MHz radial confinement.
    }
    \label{fig_MMtrends}
\end{figure}

Another important characteristic of an ion trap is its excess micromotion (EMM) profile.
In the GEN2 trap, we characterize the EMM with a single $^{\rm{171}}$Yb$^{+}$ ion using the photon-correlation technique \cite{keller2015precise} over the $^2S_{1/2} -\ ^2P_{1/2}$ transition at 370 nm, where the linewidth ($\Gamma/2\pi=19.6$ MHz) is of the same order as the RF drive frequency ($\Omega_{\rm{RF}}/2\pi= 25.65$ MHz).
We use three non-coplanar, 370 nm, red-detuned Doppler cooling beams to probe the modulation in fluorescence, and apply DC voltages on the electrodes to compensate for the EMM along the radial and axial directions. 
While we could null the radial EMM, we observe significant EMM ($E_{\rm{RF}}\sim$ 450 V/m, red circles in Fig.~\ref{fig_MMtrends}) along the trap's $x$ axis, at the trap's geometric center. By shuttling a single ion along the $x$ axis, we observed the axial micromotion null at 400 $\mu$m from the trap center, in the direction of the RF electrode pads. The observed axial micromotion null is inconsistent with the FEA simulation which predicts the null at the trap center when only the trap's blades are included, but can be explained only by taking into account the whole trap. This more complete FEA simulation points to the asymmetric capacitance between the trap's RF and DC traces on one side of the trap compared to the other as the cause of an asymmetric RF pickup, which shifts the axial micromotion null towards the RF pads (see Appendix \ref{app_trapsim}).

To mitigate this problem in the next design iteration, GEN3, we implement the following: \textbf{\emph{i)}} axial symmetrization of the RF and DC electrode routing to maintain a symmetric RF pickup on the DC electrodes, \textbf{\emph{ii)}} introducing a ground plane instead of a DC electrode at the RF-DC isolation trenches to reduce RF pickup on the DC electrodes (see bottom-left panel in Fig.~\ref{fig_trapgen}e), and \textbf{\emph{iii)}} doubling the capacitance of the in-vacuum low-pass filters (LPF) per DC electrode ($C_{\rm{LPF}} =\ 2.5$ nF) to further reduce the magnitude of the RF pickup. This pickup scales as the capacitive ratio of $C_{\rm{RF, DC}}/C_{\rm{LPF}}$, where $C_{\rm{RF, DC}}\sim\ 0.15$ pF per DC electrode, resulting in a total trap capacitance of $\sim 3\,$pF. 

In both variants of the GEN3 traps, we measure a reduced EMM profile ($E_{\rm{RF}}\sim50$ V/m) along the $x$ axis (blue triangles and black squares in Fig.~\ref{fig_MMtrends}), while being able to null the radial EMM. 
Our measurements are limited to a resolution of $\sim\pm10\,\rm{V/m}$, due to photon shot noise, laser polarization fluctuations, and power instability in our 370 nm beams. We attribute the residual axial EMM asymmetry to the close proximity of the RF wires to the DC wires on one side away from the trap stack, which is consistent in both GEN3 trap setups, and is unaccounted in the FEA simulation.

\section{\label{sec_heatingrate_trends} Motional heating rates}

\begin{figure}[b!]
    \centering
    \includegraphics[width=\columnwidth]{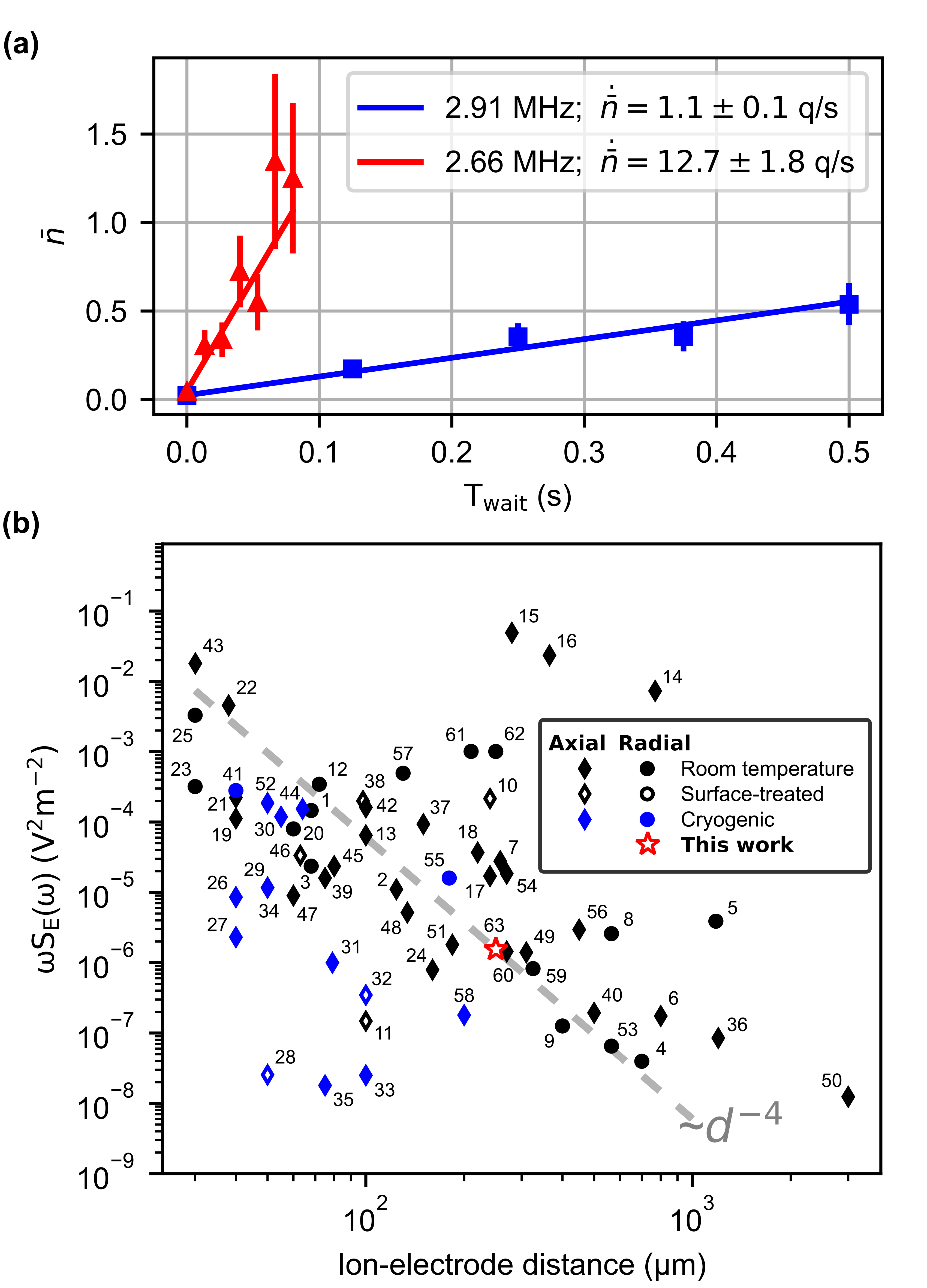}
    \caption{ 
    {\bf (a)} Heating rate of a single $^{171}\rm{Yb}^{+}$ ion extracted by measuring $\bar{n}$ using the sideband-ratio method for different wait times after sideband cooling \cite{Monroe1995a}, for the HF (blue) and LF (red) modes at $\sim480\ \rm{V}_{\rm{pk}}$ and -2 V of `twist' voltage. Error bars represent the standard deviation propagated from the measured sideband-ratio.
    {\bf (b)} A collection of surface electric field noise measurements, $S_E(\omega)$, of the axial and radial modes in multiple Paul traps as a function of the ion-electrode distance (see references in Table \ref{tab_heatingrate_legend}). The noise-spectral density is rescaled by the frequency, assuming $1/\omega$ scaling of $S_E(\omega)$. The red, open star represents $\dot{\bar n}= 1.1\pm0.1$ q/s at $\omega/2\pi= 2.91$ MHz for the GEN3-250$\mu$m trap. The dashed gray line passing through our measurement is a guiding line for the typical scaling of $d^{-4}$ from fluctuating patch potentials or dipoles on the trap electrode surface \cite{safavi2011microscopic,safavi2013influence, Brownnutt2015}. 
    }
    \label{fig_heatingrate_comparison}
\end{figure}

\begin{figure*}[t!]
    \centering
    \includegraphics[width=\textwidth]{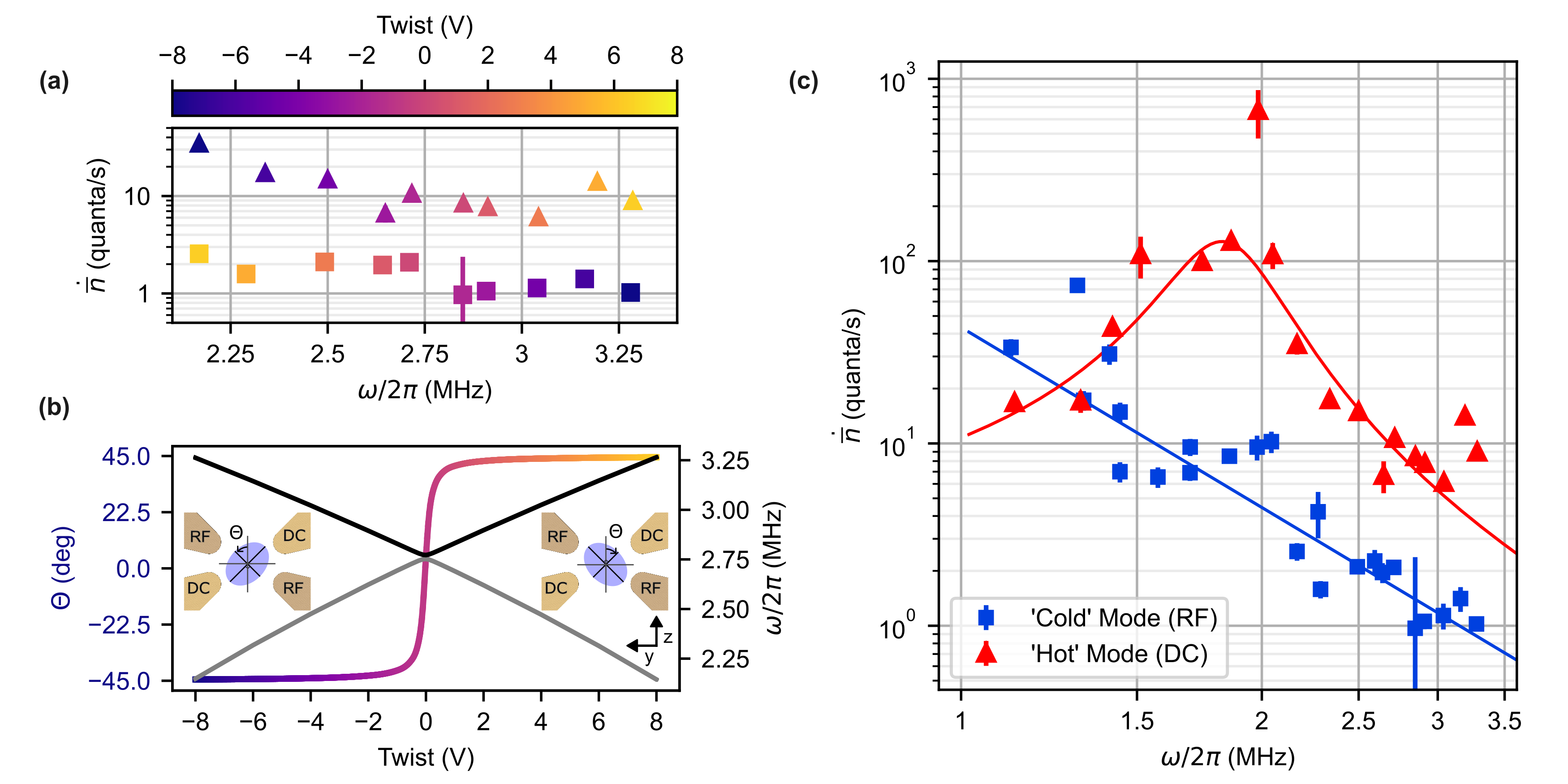}%
    \caption{ 
    {\bf(a)} Heating rate measurements at $V_{\rm{RF}}\sim480\ V_{\rm{pk}}$ for the `cold' radial mode (squares) and `hot' radial mode (triangles) as a function of radial frequency by changing the `twist' voltage (magnitude indicated by the colorbar). For a given `twist' voltage and heating rate per mode, the heating rates for each radial mode frequency can be swapped by inverting the sign of the `twist' voltage. 
    {\bf(b)} FEA simulation of the HF mode's projection angle ($\Theta$) with respect to the $z$ axis and radial mode frequency (HF-black; LF-gray) as a function of `twist' voltage. A 3 V endcap voltage lifts the radial mode degeneracy for $V_{\rm{twist}}=0\,\rm{V}$, without rotating the principal axes. The blue ellipse, between the blades, denotes an equipotential contour of the radial confinement where the minor axis is the HF mode and the major axis is the LF mode. Due to the blade angle of $\sim30^\circ$ with respect to the $y$ axis, a strong positive (negative) twist rotates the LF (HF) mode to become the `cold' mode, while the HF (LF) mode becomes the `hot' mode. 
    {\bf(c)} Heating rate of both `cold' (blue squares) and `hot' (red triangles) radial modes measured as a function of trap frequency by changing both `twist' and RF voltage. Solid blue (red) line is a fit to the `cold' (`hot') mode. Error bars represent the standard deviation in the heating rate slope extracted from a linear fit.
 }
    \label{fig_heatingratetrends}
\end{figure*}

Fluctuating electric fields originating from external electronics or electrode surfaces can excite an ion's motional mode to higher energies \cite{turchette2000heating, Deslauriers2006, Brownnutt2015}. Reducing this heating rate is often a technical challenge, particularly for the center-of-mass (COM) mode, since noisy, long-wavelength electric fields can couple strongly to its spatially homogeneous mode vector. The COM mode is a crucial resource to engineer long-range spin-spin interactions \cite{monroe2021programmable, Defenu2023, kyprianidis2024interaction, De2025} and to realize programmable spin-phonon couplings \cite{Davoudi2021, than2025observationquantumfieldtheorydynamicsspinphonon, Kang2024, pagano2025varenna}. Low heating rates and long motional coherence times benefit the above, and many other, applications, especially given that the COM heating rate scales with the system size \cite{kalincev2021motional}.

We measure the single-ion heating rates in the GEN2 and GEN3-250$\mu$m trap with a \Ybf ion \cite{Monroe1995a, leibfried2003quantum}. The GEN2 trap exhibited a heating rate of 675 q/s at 2.95 MHz for the HF radial mode, and a larger heating rate $\sim10,000$ q/s at 2.22 MHz for the LF radial mode (see Fig. \ref{fig_GEN2heatingrates} in Appendix \ref{app_heatingratetechnical}).
Conversely, in the GEN3-250$\mu$m trap, we observe a significant reduction in the heating rate of both radial COM modes of a single ion: for example, in the voltage configuration reported in Fig.~\ref{fig_heatingrate_comparison}a, we observe 1.1$\pm$0.1 q/s at 2.91 MHz (HF) and 12.7$\pm$1.8 q/s at 2.66 MHz (LF). Compared to the heating rates of both microfabricated 2D chip traps and macroscopic 3D traps (microfabricated or not), the GEN3-250$\mu$m trap exhibits one of the lowest reported heating rates around  $d=250\,\mu$m, on par with cryogenic traps despite being near room temperature (see Fig.~\ref{fig_heatingrate_comparison}b).

The main differences between the two design generations are as follows: unlike GEN2, the GEN3-250$\mu$m trap underwent \emph{ex situ} plasma cleaning. We plasma-cleaned the full trap assembly for a total of 12 min with argon and oxygen (see Appendix \ref{app_plasmacleaning}). We inserted the trap assembly into the vacuum chamber in a Class 1000 cleanroom (ISO 6) and started the vacuum pump-down 11 hours after plasma cleaning. Similarly to GEN2, we then baked the chamber for 4 weeks at $\rm{\sim180^{\circ}C}$. Other significant changes, as mentioned in Sec. \ref{sec_trapfreqMM}, include more symmetric electrical routing and larger filter capacitors to mitigate asymmetric RF pickup, as well as the presence of a ground surface on the trap.

\begin{figure*}[t!]
    \centering
    \includegraphics[width=\textwidth]{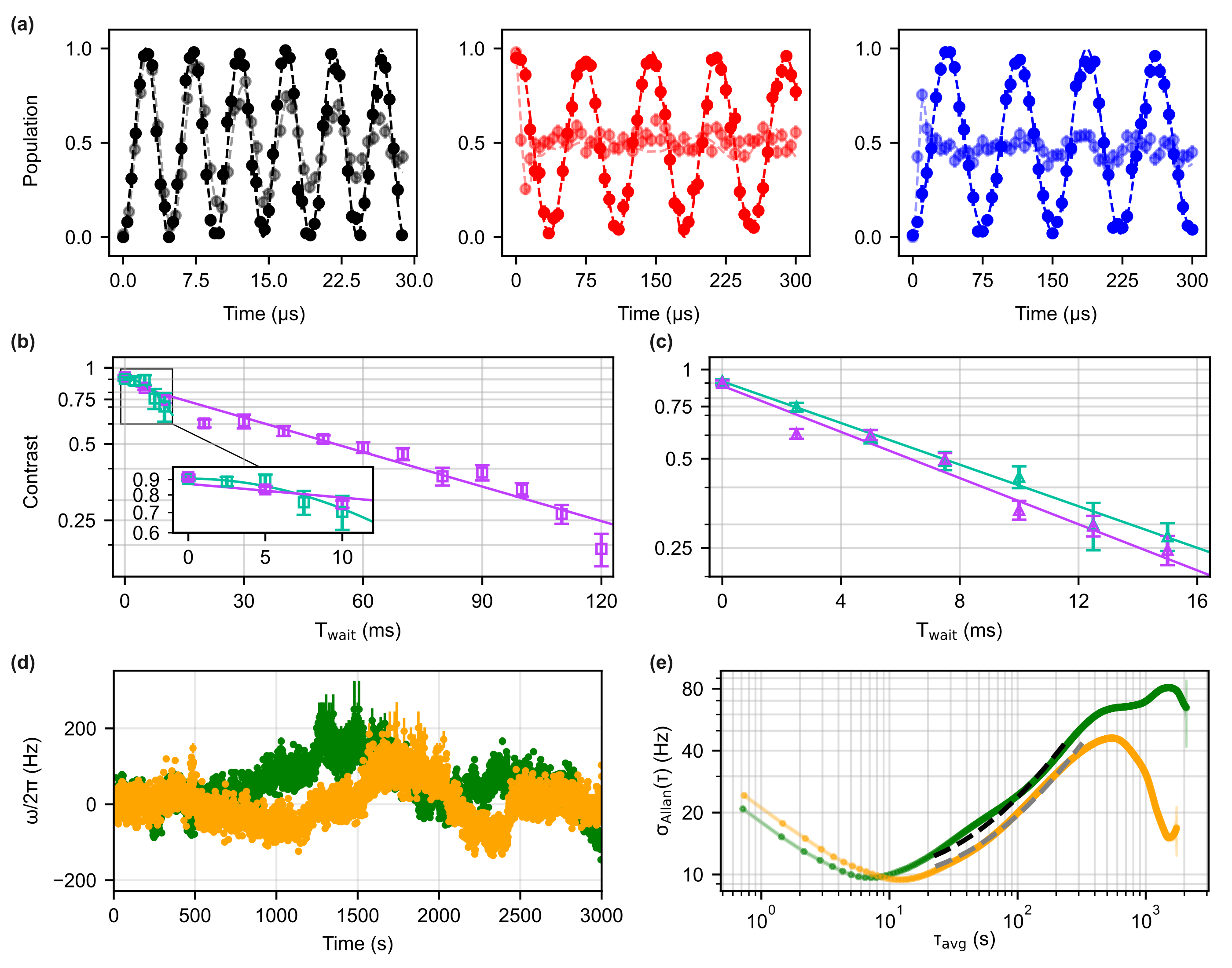}
    \caption{
    {\bf (a)} Coherent Rabi oscillations on the ground state hyperfine qubit of $^{171}\rm{Yb}^{+}$ with $\Omega_{\rm{carrier}}/2\pi\approx200 \ \rm{kHz}$ (left), HF red-sideband (center, red) with Lamb-Dicke parameter $\eta\approx0.07$, and HF blue-sideband (right, blue) after ground state cooling. The translucent (solid) data show the Rabi oscillations after Doppler cooling with $\bar n\approx11$ (SBC with $\bar n=0.06(1)$) quanta. Fits (dashed lines) to the Rabi oscillations assume a thermal phonon distribution.
    Error bars (close to marker size) represent the standard deviation of the mean thresholded counts. 
    {\bf (b),(c)} Contrast decay of a motional Ramsey phase scan for each radial mode over a variable interrogation time. In both panels, the turquoise plots have no motional decoupling, while the purple plots involve a single motional echo pulse. The decaying contrast is fit (lines) to a Gaussian function ({(b)} turquoise) for the HF (squares) mode without the motional echo, and an exponential for the other datasets, including the LF mode (triangles). Error bars are the standard deviation propagated from the sinusoidal fit of the Ramsey phase scan.
    {\bf (d)} Fluctuations in the trap frequency measured by parking at the side-of-slope of the low-power, blue-sideband, Rabi lineshape of the HF mode (yellow), and the LF mode (green), with drifts spanning $\sim$200 Hz. Error bars indicate the standard deviation in the estimated frequency by propagating the error from the measured population at the Rabi lineshape's side-of-slope. 
    {\bf (e)} Allan deviation extracted for the corresponding mode data in (d) over a varying time window ($\tau_{\rm{avg}}$). Error bands of the corresponding color represent the standard deviation in the Allan deviation estimates.  A linear fit to the slope of the data after 10 s suggests a frequency drift of $\sim$0.11 Hz/s (dashed gray) for the HF mode, and $\sim$0.15 Hz/s (dashed black) for the LF mode.
    }
    \label{fig_coherentdynamics1}
\end{figure*}

The dependence of the heating rate on the trap frequency for both radial modes suggests that technical noise may still be the main driver of the observed heating rates. Interestingly, we observe an anisotropy in the heating rate depending on the radial projection of the motional modes towards the RF or DC blades (see Fig.~\ref{fig_heatingratetrends}a,b), controlled by the `twist' voltage ($V_{\rm twist}$), which separates the two radial mode frequencies while also rotating the trap's radial principal axes. This suggests a strong anisotropy in the spatial noise profile. The radial motional mode, which increasingly projects towards the RF blades (the `cold' mode), consistently exhibits a lower heating rate than the radial mode with a stronger projection towards the DC blades (the `hot' mode). Flipping the sign of the `twist' voltage inverts the projection of the radial modes and hence exchanges their heating rate characteristics as well. By tuning $V_{\rm{twist}}$ (see Fig.~\ref{fig_heatingratetrends}a), we can configure $\dot{\bar n}<10\,\rm{q/s}$ for both radial modes.

To investigate this further, we study the dependence of the heating rate on the radial mode frequency for both the `hot' and `cold' modes by varying the RF voltage applied to the trap and the `twist' voltage. 
The saturation of the radial mode projection towards either the RF or DC blade for $|\rm{V}_{\rm{twist}}| > 1\ \rm{V}$ (our operating regime) (see Fig.~\ref{fig_heatingratetrends}b) allows us to use the `twist' as a parameter to tune the radial trap frequencies for this study, while minimally changing the mode projection angle.
As shown in Fig.~\ref{fig_heatingratetrends}c, we observe a peak near $\omega/2\pi\approx 1.8\ \rm{MHz}$ in the heating rate of the `hot' mode over the measured frequency range.
Conversely, the `cold' mode approximately follows a power law of $1/\omega^{3.2(2)}$. 
These observations indicate that the dominant noise source for both modes is not the flicker noise of surface fluctuating patch potentials and adatom diffusion \cite{Brownnutt2015}, where typically $\dot{\bar n}\sim1/\omega^2$. While the heating rate trend of the `hot' mode seems to be caused by technical noise, the trend in the `cold' mode still approximately follows a cubic power law, indicating perhaps a combination of fluctuating dipoles of surface adsorbates and technical noise \cite{Brownnutt2015, hite2021surface, kalincev2021motional, sedlacek2018evidence, safavi2013influence, safavi2011microscopic}. The source of this technical noise is still under investigation (see Appendix \ref{app_heatingratetechnical}).

\section{\label{sec_coherentmanip} Coherent Manipulation}

Clock spin qubits in trapped ions offer extremely long coherence times spanning from seconds to hours \cite{Wang2021single,an2022high}. However, multi-qubit and spin-phonon operations \cite{fang2023realizationscalableciraczollermultiqubit, Andrade2022, Than2025phase, than2025observationquantumfieldtheorydynamicsspinphonon} are often not limited by the spin coherence, but by the coherence time of the collective bosonic modes, which is restricted by the stability of the RF and DC voltages, and their respective heating rates. The low heating rates reported here ($\lesssim10$ q/s) at $\omega/2\pi\approx 3\ \rm{MHz}$ enable extended motional coherence times $\gtrsim$ 10 ms for both center-of-mass radial modes.

In Fig.~\ref{fig_coherentdynamics1}a, we show the Rabi evolution of the hyperfine carrier transition in the ground-state clock qubit of a single $^{171}\rm{Yb}^+$ ion using counter-propagating, pulsed, 355 nm Raman beams \cite{Hayes2010}. The red (after spin state inversion) and blue motional sidebands evolution also exhibit high contrast after motional ground state cooling to $\bar n=0.06(1) \ \rm{quanta}$ for the HF mode in the GEN3-250$\mu$m trap. The high signal-to-noise ratio observed in these oscillations is a result of using a 0.6 NA objective to achieve 99\% state preparation and measurement (SPAM) fidelity with a photo-multiplier tube (PMT) for a single ion \cite{Noek13}.

We characterize the coherence properties of the collective motional modes by measuring the motional Ramsey coherence time \cite{Fluhmann2019encoding} of each radial mode of a single ion, as shown in Fig.~ \ref{fig_coherentdynamics1}b,c. We first initialize the system in $\ket{\downarrow_{{z}}, 0}$ (in the $\sigma_{{z}}\otimes\hat{n}$ basis) after motional ground state cooling and optical pumping. Next, we implement an effective motional $\pi/2$ pulse by applying a carrier $\pi/2$ pulse, followed by a red-sideband $\pi$ pulse to prepare a $\ket{\downarrow_{\rm{z}}}(\ket{0} + \ket{1})/\sqrt{2}$ superposition state.
After a time $T_{\rm wait}$, we apply a final effective motional $\pi/2$ pulse (a red-sideband $\pi$ pulse and then a carrier $\pi/2$ pulse with variable phase) and analyze the contrast of the Ramsey fringe (sequence illustrated in Appendix \ref{app_motionalphasescans}).
For each $T_{\rm{wait}}$, we measure the contrast of the oscillations by scanning the phase of the second carrier $\pi/2$ pulse (see Fig.~\ref{fig_coherentdynamics1}b,c).

\begin{figure*}[!t]
    \centering
    \includegraphics[width=\textwidth]{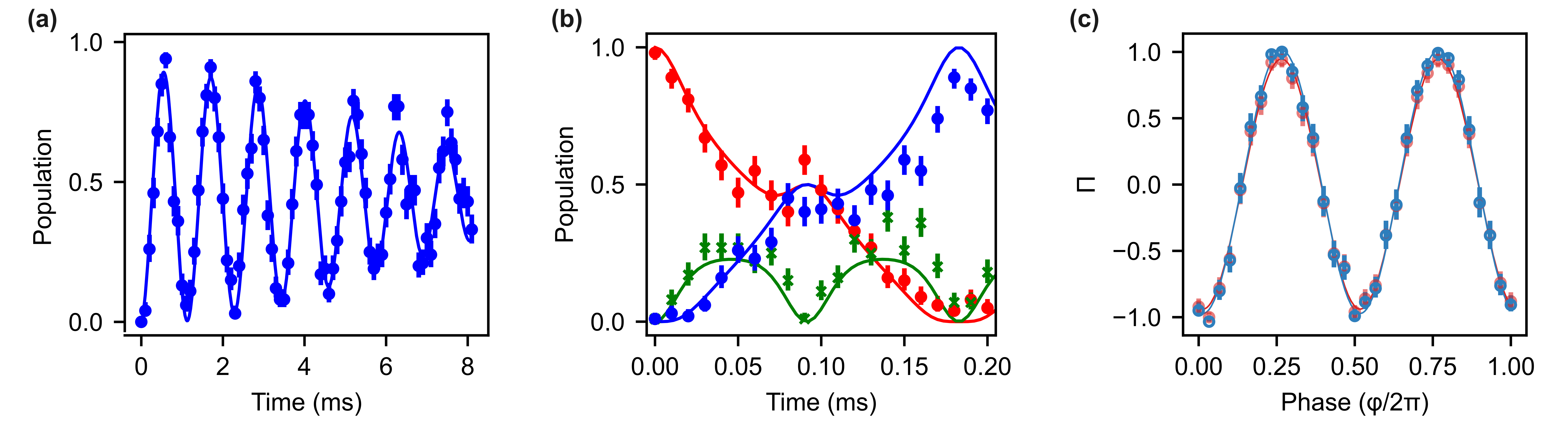}
    \caption{ 
    {\bf (a)} Ising oscillations measured from the population in $\ket{\uparrow_{{z}}\uparrow_{{z}}}$ for two ions, using a PMT, by operating in the dispersive regime of the M\o lmer-S\o rensen interaction. The coherent two-body oscillations are fit to a sinusoid, for $J_{\rm{Ising}}/2\pi=0.433(1)\,\rm{kHz}$, with a Gaussian decay envelope, $e^{-(t/7.8(4)\,\rm{ms})^2} $.
    {\bf (b)} Evolution in the resonant regime of the M\o lmer-S\o rensen interaction, measuring the population in $P_{\uparrow_{{z}}\uparrow_{{z}}}$ (blue dots), $P_{\uparrow_{\rm{z}}\downarrow_{{z}}}+P_{\downarrow_{\rm{z}}\uparrow_{{z}}}$ (green crosses), and $P_{\downarrow_{{z}}\downarrow_{{z}}}$ (red dots) states. Solid lines represent the QuTiP \cite{JOHANSSON2013} simulation of the interaction with $\eta_{im}\Omega_i/2\pi=5.47 \ \rm{kHz}$ and $\delta_m/2\pi=10.94 \ \rm{kHz}$ as used in the experiment.
    {\bf (c)} Parity oscillations after preparing the Bell state, $\ket{\Phi}$, and scanning the phase of a carrier $\pi/2$ probe pulse. Red (blue) dots are measurements without (with) SPAM correction. Each point is a statistical average of 200 repetitions.
    All the asymmetric error bars in {(a-c)} represent the 68.3\% credible interval from a Beta distribution posterior.}
    
    \label{fig_MSevol}
\end{figure*}
In the experimental configuration of this measurement, the heating rates of the HF and LF modes were 1.6(2) q/s and 48(3) q/s (limited by technical noise), respectively.
For the LF mode, we observe an exponential decay in the contrast with a time constant of $T_2^*=12.4(3)\rm{ms}$. For the HF mode, we could only satisfactorily measure the contrast up to 10 ms, after which slow drifts in the trap frequency affect the measurement during the full phase scan. This breaks down the fixed phase relationship between the local oscillator (set by the Raman beatnote) and the ion's motion that are still independently coherent over 10-50 ms. Due to the frequency noise, we fit the contrast decay to a Gaussian function and extract $T_2^*=21(3)\,\rm{ms}$, when the contrast falls to $1/e$ \cite{KiesenhoferThesis2024}.

We decoupled the slow drift from the measurement using a motional echo pulse, akin to that in Ref. \cite{mccormick2019coherently, jarlaud2021coherence}. The decoupling pulse sequence consists of a red-sideband $\pi$ pulse, a carrier $\pi$ pulse, and another red-sideband $\pi$ pulse, in sequence, to invert the motional state populations and their phase halfway across the wait time, $T_{wait}$. In this way, as shown in Fig.~\ref{fig_coherentdynamics1}b, we can refocus the phase relationship between the ion motion and the Raman beatnote and extend the motional coherence time to $T_2=96(6)\ \rm{ms}$, where the contrast decays exponentially. This decay is consistent with the measured heating rate and an additional motional dephasing of $\approx 2\pi\,\cdot\, 2.3$ Hz, obtained from numerics. Conversely, the motional echo did not improve the coherence time of the LF mode ($T_2=11.2(9)\ \rm{ms}$), which is fully explained by the heating rate. The observed motional coherence times are consistent with the predicted scaling of $\approx1/(2\dot{\bar n }\,+\,\gamma_{\text{ph}}/2)$, for a given heating rate, $\dot{\bar n}$, and motional dephasing rate, $\gamma_{\text{ph}}$ \cite{Brownnutt2015,turchette2000decoherence}. Furthermore, upon inverting the `twist' voltage, we observe an exchange in the motional coherence characteristics of the two radial modes (see Appendix \ref{app_motionalcoherence} for details)

To characterize the frequency drifts observed without the motional echo, we measure the variation in the population with side-of-fringe spectroscopy, by applying a detuned blue sideband $\pi$ pulse at low power, and calculate the Allan deviation of both the LF and HF modes (see Fig.~\ref{fig_coherentdynamics1}d,e). 
Both motional modes exhibit linear drifts of $\sim$0.15 Hz/s (LF) and $\sim$0.11 Hz/s (HF), which explains why the motional Ramsey phase scans for $>\,10$ ms still display high contrast but deviate from the expected periodicity in the absence of the motional echo pulse. Since these measurements were not taken concurrently, the differential stability between the two motional modes, separated by the twist, cannot be ascertained.

We achieve such motional frequency stability by sampling the RF voltage with an inductive, pickoff loop inside the RF resonator, which is, in turn, rectified with a circuit designed to be largely insensitive to thermal drifts \cite{Johnson2016}. This design is based on a modified bridge rectifier and reaches a similar thermal stability as Ref. \cite{Johnson2016} but provides twice the signal-to-noise ratio (see Appendix \ref{app_RFsetup}).
The RF voltage stabilization could be further improved by using a compound PID controller, for accurate compensation in the slow ($>$ 1 s) and fast (0.1 ms - 1 s) timescales, and temperature stabilization of the RF circuit elements \cite{KiesenhoferThesis2024, harty2013high,metzner2024sound}.

To evaluate the system’s capabilities for high-fidelity quantum simulation and computation, we characterize the M\o lmer-S\o rensen interaction in both the dispersive \cite{Molmer1999} ($\eta_{im}\Omega_i \ll\delta_m $) and resonant regimes ($\eta_{im}\Omega_i \sim\delta_m $)  \cite{Sorensen2000} using two $^{171}\text{Yb}^{+}$ ions separated by $4.4\,\mu\text{m}$. We observe coherent Ising oscillations after symmetrically detuning the bichromatic Raman beatnote from the motional sidebands of the HF mode, by $\delta_m/2\pi\approx+28\,\rm{kHz}$, with $\eta_{im}\Omega_i/2\pi\approx8\,\rm{kHz}$ (Fig.~ \ref{fig_MSevol}a). By tuning to $\delta_m/2\pi\approx+10\,\rm{kHz}$ with $\eta_{im}\Omega_i/2\pi\approx5\,\rm{kHz}$, we observe coherent one and two-body,  M\o lmer-S\o rensen oscillations as shown in Fig.~\ref{fig_MSevol}b, limited by fluctuations in laser power and trap frequency. With a single rectangular pulse, we prepare a Bell-state, $\ket{ \Phi}=(\ket{\downarrow_{\rm{z}}\downarrow_{{z}}}+i\ket{\uparrow_{\rm{z}}\uparrow_{{z}}})/\sqrt{2}$, and achieve a parity contrast
of $94.3^{+0.7}_{-0.8}$ \%  and $99.3^{+0.7}_{-1.5}$ \%, before and after SPAM correction, respectively, by fitting to the parity oscillation using the maximum likelihood estimation (MLE) method (see Appendix \ref{sec_technicalMS}). Amplitude and frequency-based pulse-shaping techniques may be used to further improve the fidelity \cite{jia2023angle, debnath2016demonstration}.

\section{\label{sec_discussion} Discussion}

In this work, we demonstrate the first microfabricated, monolithic segmented 3D trap that simultaneously features: \emph{(i)} resilience to high RF voltages for deep trapping potentials, \emph{(ii)} multi-directional, large NA optical access for efficient state detection and single-ion manipulation, and \emph{(iii)} motional mode performance comparable to the best macroscopic and cryogenic systems \cite{Whitlow2023, KiesenhoferThesis2024, Ballance16, matsos2025universal}. These combined characteristics will enable progress in a broad range of quantum technology applications, from computing \cite{Bruzewicz2019} and simulation \cite{monroe2021programmable} to networking \cite{Duan2010} and metrology \cite{Ludlow2014}.

For example, while previous monolithic 3D traps have been characterized with light ions such as \Ca \cite{Xu20253D, Lovera2024, Kiesenhofer2023}, the thermal and electrical performance reported here extends the high-performance capabilities to heavy ionic species, such as \Yb, \Ba, and Lu$^+$\cite{Zhang2023Lu, arnold2025opticalclocksaccuracyvalidated}.

Moreover, the motional properties of the GEN3-250$\mu$m trap at high secular frequency ($\sim1$ q/s heating rates, $T_2\sim 95$ ms motional coherence time, and high radial frequency stability) enable improved fidelity for both analog simulation and digital operations, including hybrid analog-digital simulations
\cite{Davoudi2021, than2025observationquantumfieldtheorydynamicsspinphonon, padilla2025vibrationallyassistedexcitontransfer,katz2025hybriddigitalanalogprotocolssimulating} (see Fig.~\ref{fig_MSevol}a,b). In particular, a high radial frequency helps to reduce spectral crowding of radial modes, and reduces the carrier transition's crosstalk in spin-phonon operations \cite{James1998, pagano2025varenna}. Achieving low heating rates on both radial branches enables their use for different purposes, including parallel two-qubit gates on disjoint qubit pairs \cite{figgatt2018parallel, Zhu2023Pairwise}, simulation of para-particles \cite{Alderete2025}, and simulations of out-of-equilibrium dynamics of vibronic models \cite{pagano2025varenna, so2024trappedion, so2025, so2025experimentalrealizationthermalreservoirs}. A well-behaved center-of-mass mode could be used for the efficient preparation of W-states \cite{Retzker2007Tavis, Zhu2025} and for dark matter searches \cite{Carney2021, Budker2022} that need long-interrogation times. Similarly, the demonstrated axial homogeneity and electrode segmentation enable operations with long chains \cite{Pagano2018}, where uniformity of confinement reduces calibration overhead and suppresses spatially dependent systematic errors in addressing, cooling, state readout, and gate performance.

The high, multidirectional optical access of this trap and its small footprint make it an ideal candidate for integrating cavities \cite{Schupp2021, Krutyanskiy2023, teh2024ion, gao2023optimization} and external photonic chips \cite{lim2025scalableionfluorescencecollection, craft2026low} to enhance photon-collection rates for quantum networking protocols. Additionally, the combination of multi-MHz confinement and low heating supports high-duty-cycle networking nodes. This enables long sequences of photonic attempts while remaining within the Lamb–Dicke regime without recooling. {This reduces the impact of photon-recoil-induced errors occurring in both polarization-based \cite{Stephenson2020} and time-bin-based \cite{Saha2025} remote entanglement protocols \cite{Kikura2025, Yu2026, Apolin2026}.}

The low motional heating presented here supports quantum logic spectroscopy of species that lack cycling transitions, including molecular ions \cite{Wolf2016NatureQLS, Chou2017NatureMoleculeQLS} and highly charged ions \cite{Micke2020NatureHCIQLS, King2022NatureHCIClock}. 
In these applications, any residual heating or motional dephasing directly reduces the fidelity of mapping transition probabilities across ion species by broadening motional features and introducing time-dependent Doppler and Stark shifts during long interrogations.

While the current, GEN3, design is already suitable for a single long ion chain, $N\sim20-30$, for both quantum logic and networking, the SLE technology allows for scaling to more electrodes with a finer pitch, which would enable trapping of longer, equispaced ion chains with multiple zones for dedicated operations, including loading, cooling, state detection, and computation.
 
Finally, the thermal packaging and voltage-handling strategies demonstrated here can be extended to monolithic geometries with larger aspect ratios designed to trap large 2D ion crystals \cite{Donofrio2021Radial, Wang2020coherently, Kiesenhofer2023, Qiao2024tunable, Guo2024site-resolved}. Here, using a large RF frequency drive and mitigating its power dissipation  ($P_{\rm diss}\sim\Omega_{RF}^2$) becomes especially important to reduce the effects of micromotion away from the RF null.

\begin{acknowledgments}
G.P. and A.M. thank S. De, L. Zhang, C. Coss, and Y. Chen for their contributions to the trap characterization, and J. Chiaverini for insightful discussions on plasma cleaning.
N.M.L. thanks S. Decoppet and U. Singla for their work towards trap testing. P.B. and M. D. thank Chris Schenk for his contribution to this work with the fabrication of the traps. 

G.P. acknowledges that this material is based on work supported by the U.S. Department of Energy, Office of Science, Office of Nuclear Physics under the Early Career Award (grant no. DE-SC0023806).
G.P. also acknowledges the Welch Foundation Award (grant no. C-2154), the Office of Naval Research Young Investigator Program (grant no. N00014-22-1-2282), the NSF CAREER Award (grant no. PHY-2144910), and the Office of Naval Research (grant no. N00014-23-1-2665 and N00014-24-1-2593). 
G.P., N.M.L. and P. B. acknowledge the Army Research Office (award
W911NF21P0003), and the US Army Research Laboratory (award W911QX21C0031 P00002).
P. B. acknowledges that this material is based on work supported by the U.S Department of
Energy, Office of Science (grant no. DE-SC0020553).
N.M.L. acknowledges that this material is based on work supported by the U.S Department of Energy, Office of Science, Office of Nuclear Physics under the Early Career Award (grant no. DE-SC0024504). N. M. L. acknowledges support from the National Science Foundation (QLCI grant OMA-2120757).

\end{acknowledgments}

\appendix

\section{\label{app_projectworkflow} Project workflow}

Our collaborative project between groups at Rice University and Duke University, and Translume Inc. has the following workflow. First, the monolithic trap is designed by the collaboration, which involves optimizing the trap design for desired characteristics using FEA, while remaining within the constraints of the SLE technique. 
Translume Inc. fabricates the trap using their microfabrication processes and verifies the basic electrical connectivity, prior to sending the packaged traps to the groups at the universities, who test multiple traps of different generations under vacuum. At each university, we first characterize the macroscopic thermal characteristics of the trap assembly, and then measure microscopic properties of the trapping potential and implement quantum operations using trapped ions as the probe. 

\begin{figure}[t!]
    \centering
    \includegraphics[width=\columnwidth]{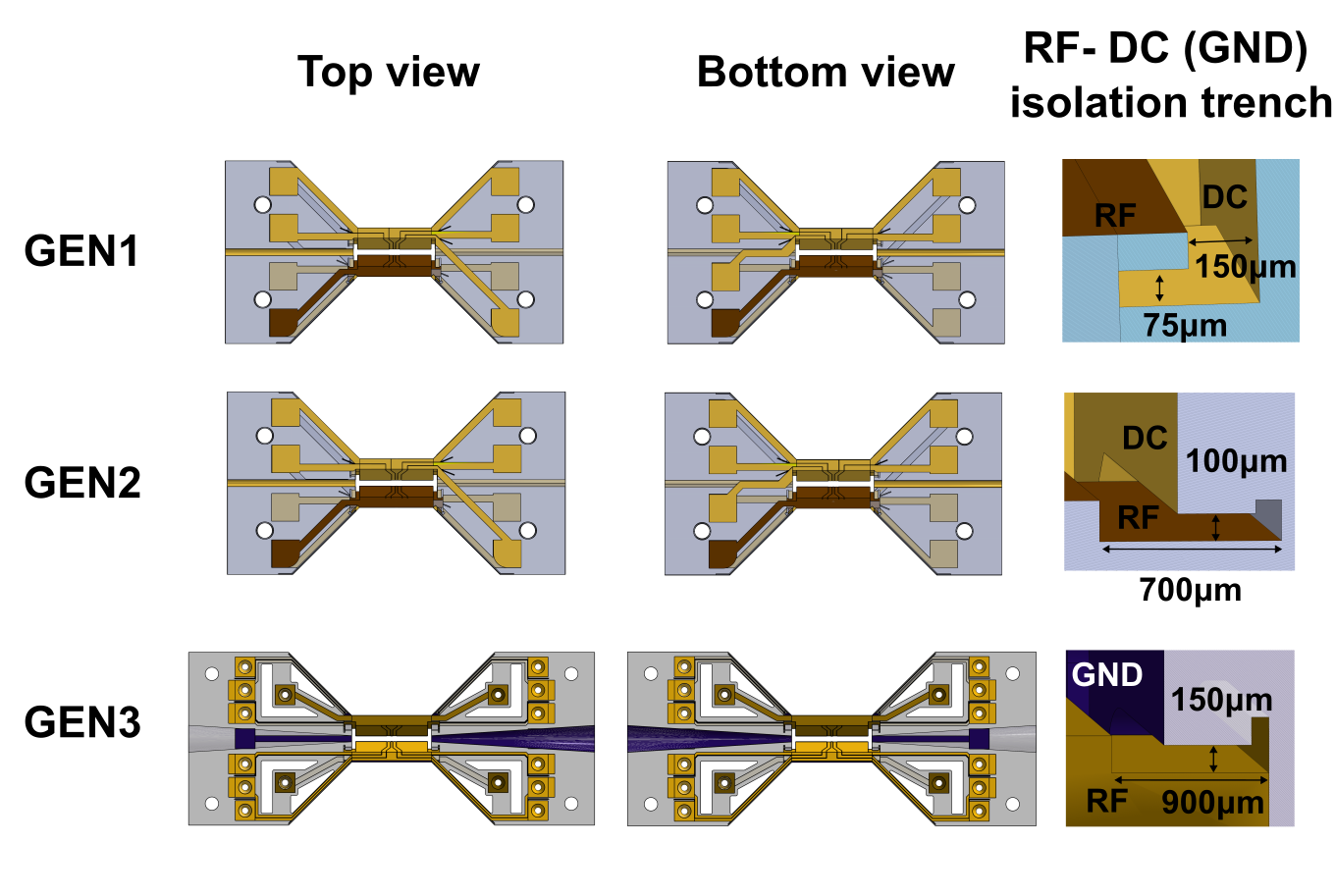}
    \caption{ 
    Designs of each trap generation with modifications in trap geometry and electrical routing, including the RF-DC (Ground) isolation trenches. The top and bottom views are rotated 180$^\circ$ with respect to each other about the trap's $x$ axis (axial direction).
    }
    \label{fig_trapgendevelop}
\end{figure}

Our workflow is iterated to improve the properties of each trap generation to reach as close as possible to our objective for the trap, as defined in Sec. \ref{sec_intro}. We start with GEN1 as our prototype (see Fig.~\ref{fig_trapgendevelop}). GEN2 improves on GEN1 with wider RF-DC trenches, for reasons mentioned in Sec.~\ref{sec_designandassembly}, and has rounded blade tips. Both variants of GEN3 mostly preserve GEN2's trench geometry, introduce symmetrically routed electrodes with a ground plane opposing the RF electrodes, improve thermal dissipation, and undergo \emph{ex situ} plasma cleaning before the traps are tested in vacuum.

\begin{table*}
\caption{Comparison of mechanical, electrical, and thermal properties for common ion trap substrate materials. Values are compiled for standard room-temperature operation.}
\label{tab:trap_materials}

\begin{ruledtabular}

\begin{tabular}{cccccccc}

\textbf{Property} & \textbf{AlN} & \textbf{Alumina}\footnote{99.5\% purity (AD-995 or equivalent).} & \textbf{Macor} & \textbf{Diamond} & \textbf{Sapphire} & \textbf{Silicon} & \textbf{Fused Silica} \\
\midrule

\raggedright Young's modulus (GPa) & 310--340\textsuperscript{\cite{accuratus2024}} & 380\textsuperscript{\cite{coorsTek2023}} & 67\textsuperscript{\cite{corningMacor2021}} & 1000--1200\textsuperscript{\cite{crcHandbook2022}} & 345--430\textsuperscript{\cite{sapphire2009}} & 130--190\textsuperscript{\cite{nistSilicon2023}} & 70--74\textsuperscript{\cite{accuratus2024,TranslumeWebsite}} \\
\addlinespace
\raggedright \parbox{3.5cm}{Ionization Threshold/\\ Work Function (eV)} & 6.0--7.5\textsuperscript{\cite{accuratus2024}} & 8.5--9.0\textsuperscript{\cite{coorsTek2023}} & 8.0--9.0\textsuperscript{\cite{corningMacor2021}} & 5.45--5.6 \textsuperscript{\cite{crcHandbook2022}} & 9.9\textsuperscript{\cite{sapphire2009}} & 4.6--4.9\textsuperscript{\cite{nistSilicon2023}} & 8.0--9.0\textsuperscript{\cite{accuratus2024}} \\
\addlinespace
\raggedright Dielectric Constant, $\epsilon_r$ & 8.5--9.5\textsuperscript{\cite{accuratus2024}} & 9.8\textsuperscript{\cite{coorsTek2023}} & 5.67\textsuperscript{\cite{corningMacor2021}} & 5.5--5.7\textsuperscript{\cite{crcHandbook2022}} & 9.3--11.5\textsuperscript{\cite{sapphire2009}} & 11.7--12.1\textsuperscript{\cite{nistSilicon2023}} & 3.7--3.9\textsuperscript{\cite{accuratus2024,TranslumeWebsite}} \\
\addlinespace
\raggedright  \parbox{3.5cm}{Dielectric Strength \\(kV/mm)} & 12--20\textsuperscript{\cite{accuratus2024}} & 18.5\textsuperscript{\cite{coorsTek2023}} & 40\textsuperscript{\cite{corningMacor2021}} & 500--2000\textsuperscript{\cite{crcHandbook2022}} & 20--50\textsuperscript{\cite{sapphire2009}} & 30--100\textsuperscript{\cite{nistSilicon2023}} & 20--40\textsuperscript{\cite{accuratus2024}} \\
\addlinespace
\raggedright VSF (kV/mm) & 8--18\textsuperscript{\cite{miller1993VSF}} & 8--15\textsuperscript{\cite{miller1993VSF}} & 8--12\textsuperscript{\cite{miller1993VSF}} & 20--50\textsuperscript{\cite{miller1993VSF}} & 10--25\textsuperscript{\cite{miller1993VSF}} & 2--5\textsuperscript{\cite{miller1993VSF}} & 5--12\textsuperscript{\cite{miller1993VSF}} \\
\addlinespace
\raggedright  \parbox{3.5cm}{Tangent Loss,\\ $\tan\delta$ ($\times 10^{-4}$) @ 1 MHz} & 5--15\textsuperscript{\cite{accuratus2024}} & 1--4\textsuperscript{\cite{coorsTek2023}} & 50\textsuperscript{\cite{corningMacor2021}} & 0.1--10\textsuperscript{\cite{crcHandbook2022}} & 0.1--0.5\textsuperscript{\cite{sapphire2009}} & 1--100\textsuperscript{\cite{nistSilicon2023}} & 0.1--1\textsuperscript{\cite{accuratus2024,TranslumeWebsite}} \\
\addlinespace
\raggedright  \parbox{3.5cm}{Thermal Conductivity \\ (W/m$\cdot$K)} & 140--230\textsuperscript{\cite{harris1991}} & 35\textsuperscript{\cite{coorsTek2023}} & 1.46\textsuperscript{\cite{corningMacor2021}} & 1000--2200\textsuperscript{\cite{crcHandbook2022}} & 33--42\textsuperscript{\cite{sapphire2009}} & 120--160\textsuperscript{\cite{nistSilicon2023}} & 1.3--1.4\textsuperscript{\cite{accuratus2024,TranslumeWebsite}} \\
\addlinespace
\raggedright  \parbox{3.5cm}{Thermal Expansion \\ Coefficient (ppm/K)} & 4.0--4.8\textsuperscript{\cite{harris1991}} & 8.2\textsuperscript{\cite{coorsTek2023}} & 9.3\textsuperscript{\cite{corningMacor2021}} & 0.8--1.5\textsuperscript{\cite{crcHandbook2022}} & 5.0--7.7\textsuperscript{\cite{sapphire2009}} & 2.3--2.7\textsuperscript{\cite{nistSilicon2023}} & 0.4--0.6\textsuperscript{\cite{accuratus2024,TranslumeWebsite}} \\
\addlinespace

\bottomrule
\end{tabular}
\end{ruledtabular}

\end{table*}

\section{\label{app_thermtestdata}Thermal management}

\begin{figure*}[t!]
    \centering
    \includegraphics[width=\textwidth]{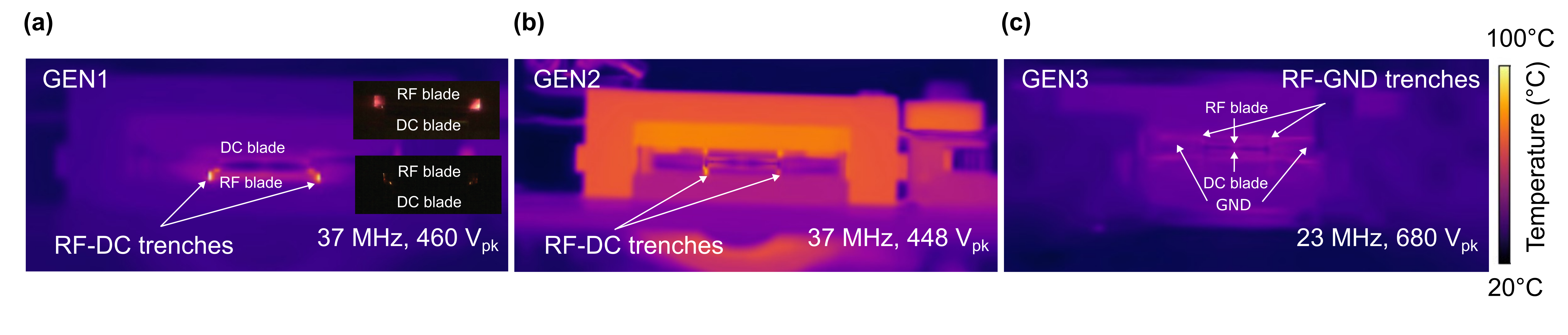}
    \caption{ 
   Thermal images of the  {\bf (a)} GEN1, {\bf (b)} GEN2 and {\bf (c)} GEN3-250$\mu$m trap taken with approximately the same RF power, using the measurement scheme described here. Insets in {\bf (a)} view the trap with an exposure time of 4 seconds from the opposite side, but in the visible spectrum. These indicate the reduction (but not elimination) of the visible glowing in the RF-DC trenches from the top to the bottom panel for similar RF power, after improved cleaning practices.  This visible glowing was absent in the successive trap generations ((b), (c)). The temperature scale is calibrated with respect to the emissivity of the AlN ceramics, for which the reading is most accurate. 
    }
    \label{fig_thermaltests}
\end{figure*}

Thermal management of ion traps is critical to ensure sufficient dissipation of high-voltage RF power away from the trap and to avoid hot spots in high vacuum conditions. Given the low thermal conductivity of fused silica, the monolithic trap's thermal interface is all the more important. Our test setup for imaging the heat distribution consists of imaging the trap assembly from outside the vacuum chamber ($\lesssim10^{-7}$ Torr) using a FOTRIC 346A-L25 IR camera through a ZnSe window on one side \cite{Hainzer2023, nordmann2020subKTraptempstabiliz}. On the opposite side, we simultaneously imaged the trap with a microscope through a fused-silica window. While driving the RF blades of the trap at high power, we monitored the RF pickoff from our quarter-wave resonator to assess the stability of the RF circuit Q-factor (See Appendix  \ref{app_RFsetup}).

We expect this system to follow a first-order differential rate equation given by the heat equation \cite{hahn2012heat}, therefore, in thermal equilibrium, T$_{\rm{local}}$ $\propto$ V$^2_{\rm{pk}}$, where T$_{\rm{local}}$ is the temperature of a local region in the assembly. The proportionality constant of this relation contains information about the thermal and electrical properties of the trap circuitry. Our aim was to minimize the rate of temperature increase and maximize its uniformity across the entire trap assembly as RF power increased, thus reducing the number of hotspots in the design.

Although we observed linearity between T$_{\rm{local}}$ and V$^2_{\rm{pk}}$ for GEN1, we observed significant heating at the RF-DC isolation trenches, along with visible radiation upon approaching 400 Vpk ($\Omega_{\rm{RF}}/2\pi= 36.6\ \rm{MHz} $), as shown in Fig.~\ref{fig_thermaltests}a. We consistently observed a steady degradation in the RF circuit's quality factor, indicated by a diminishing pickoff voltage, as the applied voltage approached 400-500 V$_{\rm{pk}}$. This was verified in six GEN1 traps at Rice and Duke University, prompting a necessary thermal design improvement prior to ion trapping. Precisely quantifying the thermal heating rate of the GEN1 trap is challenging. The IR images indicated $\sim100^{\circ}$C while the visible yellow-orange emission of the RF-DC trenches, assuming black-body radiation with the low emissivity of polished gold ($0.02-0.05$), indicated $> 500^\circ$C. Our pre-test room-temperature calibration, limited spectral sensitivity ($\lambda\colon$7-14$\mu$m), and limited spatial resolution of the IR imaging were insufficient to reliably measure such locally high temperatures. 

The observed spectral discrepancy in the imaging could point to a secondary phenomenon in addition to simple ohmic heating. We attribute this localized heating to the concentrated charge density and RF current at the gold-coated edges of the trenches, leading to high surface electric fields. Such high electric fields, particularly in vacuum, could cause VSF to promote electron emission and avalanche processes across electrodes accompanied by plasma discharge and local metal melting in the visible spectrum. These phenomena were observed previously at the `triple-junction' between the metal, vacuum, and dielectric \cite{zhang2025global, kumar2025vacuum, pillai1985surface}. The insufficient heat dissipation at the RF-DC (or Ground) junctions further contributes to the RF heating. 

The root cause of our observations was difficult to validate due to the limited IR and visible imaging resolution of the trenches in the vacuum chamber. Nevertheless, both IR and visible data pointed to the necessity of increasing the RF-DC trench gap, where we define the insulating gap length as $d_{\rm ins}$. This would reduce the local capacitance and hence ohmic heating guided by Eq.~\eqref{eqn:RF dissip} where $P_{\rm dissip}\propto1/d_{\rm ins}^2$, and minimize the affinity for VSF, where the VSF breakdown voltage $V_{\rm VSF}\propto\sqrt{d_{\rm ins}}$ \cite{pillai1985surface}.

In subsequent generations, we increased the shortest isolating dimension of the RF-DC trenches from 75 $\mu$m to 150 $\mu$m (see Fig. \ref{fig_trapgen}). We performed a similar thermal test for the GEN2 and GEN3 ion traps. We observed significantly less heating without any visible radiation up to our maximum test voltage of 1000 V$_{\rm{pk}}$ at 37 MHz for GEN2 and 900 V$_{\rm{pk}}$ at 23 MHz for GEN3-250$\mu$m. Similar results were observed in the GEN3-200$\mu$m trap. The IR camera's calibration for GEN2 was verified with an independent measurement of heating the trap with a laboratory hot-plate outside vacuum with similar imaging resolution as during the RF test, and that for GEN3-250$\mu$m was verified with a thermocouple firmly in contact with the aluminum nitride ceramics of the trap assembly in vacuum. Furthermore, upgrading the ceramic structure in the GEN3 design's trap stack with an aluminum heat sink to channel heat from the top to the bottom aluminum base plate helped evenly distribute the RF heat load, as shown in Fig.~\ref{fig_thermaltests}.
The results of the thermal tests in GEN2 and GEN3 enabled the microscopic characterization of the trap by trapping Yb$^{+}$ ions.

Despite the trap operating above room-temperature at a radial frequency $\sim3$ MHz (483 $\rm{V}_{\rm{pk}}$, $\Omega_{\rm{RF}}/2\pi=23.24\,\rm{MHz}$ with Mathieu stability parameter, $q\approx0.35$), we observe good ion-chain lifetimes with an ion-gauge reading of $\sim\,9\cdot 10^{-12}$ Torr (calibrated to N$_2$). A single ion can remain in the trap for days with Doppler cooling. A multi-ion crystal ($N=2-21$) requires recrystallization by lowering and raising the trap confinement, on average every 20-30 minutes, after melting events induced by collisions with the background gas.
These collision-induced melting events are observed in many room-temperature trapped-ion systems and typically lead to the loss of the entire ion crystal. While we have not conducted a systematic study on the ion lifetime, we speculate that our ability to recover the multi-ion crystal ($N= 2-21$) most of the time after a melting event \cite{Langevin1905, Alheit1996, Drakoudis2006} is due to the high degree of harmonicity of the eV-deep trap potential (see Appendix \ref{app_trapsim}) and low heating rates (see Sec. \ref{sec_heatingrate_trends}).

\section{\label{app_plasmacleaning} Plasma cleaning}

We implemented plasma cleaning on the GEN3 traps to reduce the heating rate compared to the GEN2 trap. Motivated by the results of \emph{ex situ} ion-milling in Ref. \cite{sedlacek2018evidence}, we tested an \emph{ex situ} plasma cleaning recipe to avoid modifying our vacuum chamber for \emph{in situ} surface cleaning techniques \cite{mcconnell2015reduction, daniilidis2014surface, mckay2014iontrapelectrodepreparationne}. The plasma cleaning process depends on the type of gases, gas flow rate, pressure, volume of chamber, RF power, and plasma cleaning time. The machine we used was the Plasma Etch PE-100 in a shared cleanroom facility. Our recipe typically consists of one cycle of 150 W of RF power, 50 sccm (standard cubic centimeter per minute) of Ar and 15 sccm of O$_{2}$ for 30 s, and another cycle of 50 sccm of Ar for 30 s at the same RF power. The average pressure in the chamber was around 50-100 mTorr during the cleaning. 

The safe time and power range were inferred by cleaning a sacrificial, older-generation monolithic trap for a long time and applying high RF power (250 W for 20 min), ensuring that no shorts developed and that no visible surface damage was observed on the trap. The GEN3-250$\mu$m trap underwent six rounds of plasma cleaning. Each round consisted of cleaning with our recipe, rotating the trap by 180$^\circ$ about its $x$ axis, and then cleaning again. The first four rounds were performed with the trap alone, and the last two were with the entire trap assembly. The trap was also exposed to air for hours between these sessions. The trap electrodes were shorted to the plasma cleaner's ground during the process to avoid charging. After the last cleaning stage, the trap was exposed to air for 11 hours during vacuum assembly before the vacuum-pump-down. Despite baking the system, we still avail some of the benefits of plasma cleaning to observe $\dot{\bar n}\sim 1\,\rm{q/s}$, which perhaps may not be as much as \emph{in situ} surface treatment after the bake \cite{mckay2014iontrapelectrodepreparationne, hite2012100}. 

\section{\label{app_MMC_drifts} Micromotion and Drifts}

\begin{figure}[b!]
    \centering
    \includegraphics[width=\columnwidth]{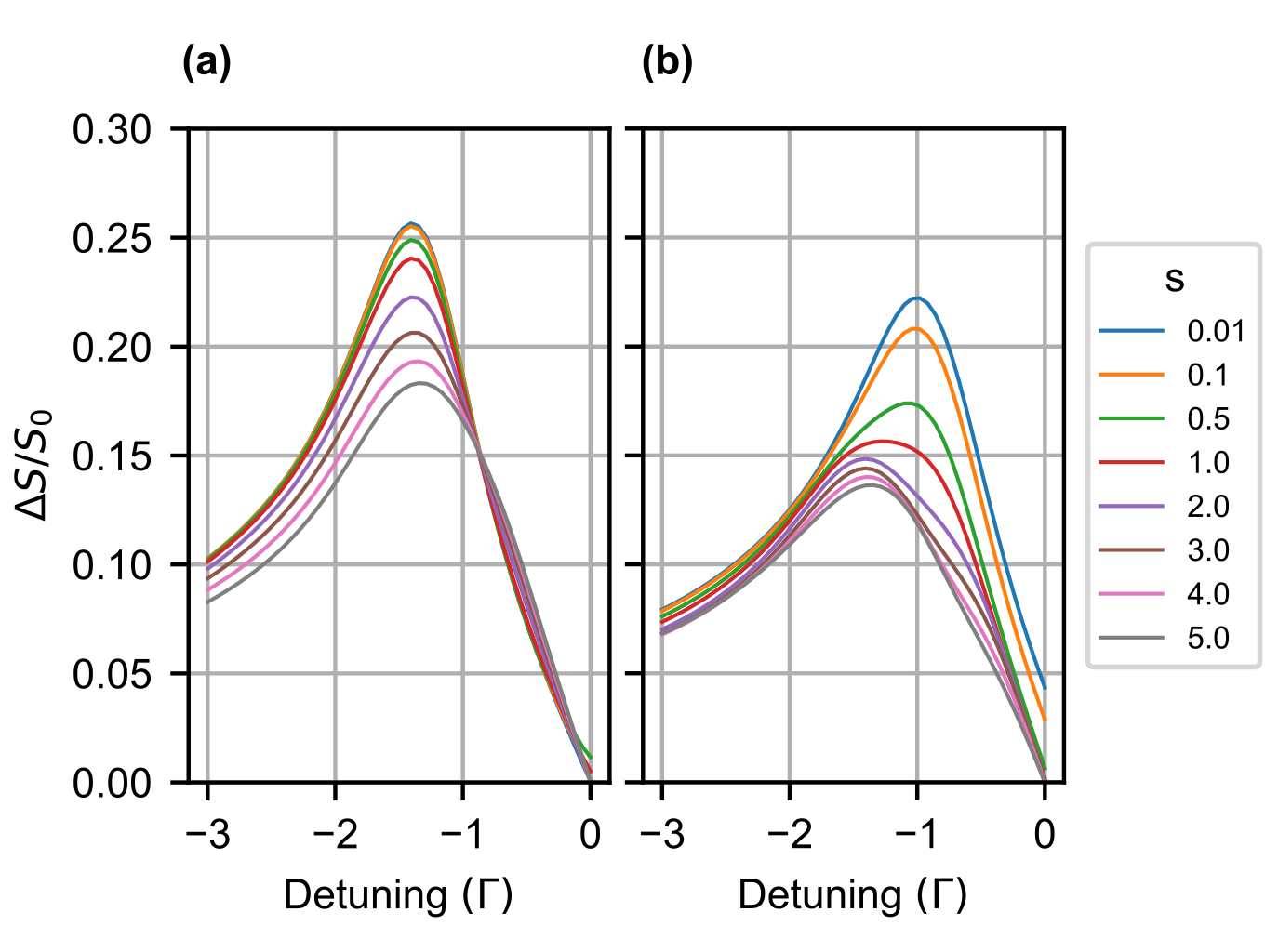}
    \caption{ 
    Numerical simulation of the contrast, $\Delta S/S_0$, of the modulation in fluorescence due to excess micromotion in {\bf (a)} \Ybf and {\bf (b)} \Ybc for different saturation parameters, $s$, as a function of detuning of the Doppler cooling beam from resonance. Here the modulation strength $\beta=0.1$, and magnetic field strength $B=4.5$ Gauss. The RF frequency is $\Omega_{\rm{RF}}/2\pi=25.6$ MHz, 23.2 MHz, and 22.9 MHz while characterizing the (a) GEN2, (b) GEN3-250$\mu$m, and GEN3-200$\mu$m trap.
    }
    \label{fig_MMsimulations}
\end{figure}

We implement the photon-correlation technique \cite{keller2015precise} to characterize the EMM along the radial and axial directions of the GEN2 and GEN3 traps. To accurately model the steady state modulation of fluorescence in \Ybf (used in GEN2 and GEN3-200$\mu$m) and \Ybc (used in GEN3-250$\mu$m), we numerically simulate a four-level optical Bloch equation (OBE) of the cooling transition in \Ybf and \Ybc, as shown in Fig.~\ref{fig_MMsimulations}. We simulate the OBE with the parameters used in the experiment (saturation parameter, magnetic field strength, polarization etc.) for different modulation strengths, $\beta$. From this, we extract the steady state contrast of the oscillation in the excited state population: the ratio of amplitude of modulation in population over the  average population, $\Delta S/S_0$. We obtain a monotonic relation between $\beta$ and $\Delta S/S_0$ after interpolating the relation for a finite number of simulations. By fitting the contrast measured in the experiment over a RF period to a sinusoid, we determine $\beta$ and $E_{\rm {RF}}= \beta m\Omega_{\rm RF}^2/ k\,e$, where $e$ is the ion's charge, $m$ is the mass of the ion, and $k$ is the wavevector of the micromotion compensation beam.

While the EMM compensation of the GEN2 trap required along the $y$ and $x$ directions did not drift, the radial EMM compensation along $z$ drifted erratically ($E_{\rm RF}$ along $y$ axis) requiring compensation voltages to be adjusted up to 20 mV ($z$ direction `push') roughly every 5 min at $\omega/2\pi\sim 2\,\rm{MHz}$, and 50 mV every 5 min at $\omega/2\pi\sim 3\,\rm{MHz}$ accompanied by $\pm2\mu$m axial excursions in minutes along with a continuous drift in position. These instabilities rendered the trap unsuitable for continuous stable operation in this configuration.

The improved design in the GEN3-250$\mu$m trap assembly reduced  the drifts in the EMM along the radial directions to the level of  $\pm 10\,\rm{mV}$ in compensation (corresponding to $\sim\pm80$ nm displacement from the RF null) at $\omega/2\pi\sim 3\ \rm{MHz}$, which tends to be more prominent along the vertical $z$ direction. Both radial EMM drifts and axial positional excursions, $\pm\,0.5\,\mu$m at low axial confinement ($\sim320$ kHz), occur on timescales of hours, which makes both effects manageable with periodic calibration routines.

To further reduce the effects of EMM and positional drifts at high RF voltages while maintaining high radial frequencies, we design a variant of the GEN3 trap with $d=200\,\mu$m, referred to as GEN3-200$\mu$m in the text. In Fig.~\ref{fig_MMtrends}, we report the characterization of its axial micromotion profile, which stays below 100 V/m over a wide range of axial positions similar to the GEN3-250$\mu$m trap. In the GEN3-200$\mu$m trap, we also observe improved positional drifts $\sim\pm0.25\,\mu\rm{m}$ over several hours.

\section{Motional mode characterization}

\subsection{\label{app_radialfreqfits} Radial frequency measurements}

\begin{figure}[b!]
    \centering
    \includegraphics[width=\columnwidth]{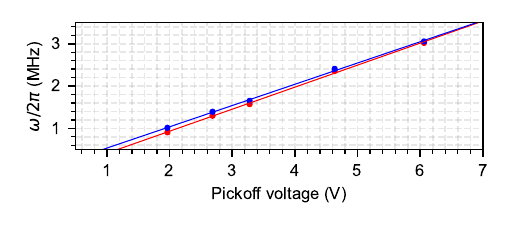}
    \caption{ 
    Linear dependence of the radial frequency versus applied RF voltage, measured by sampling a fraction of the applied voltage called the `pickoff voltage'. The slope of the HF mode (blue) is 0.50(1) MHz/$V_{\text{pickoff}}$, and that of the LF mode (red) is 0.52(1) MHz/$V_{\text{pickoff}}$ , for -0.3 V `twist' voltage. These measurements, using `RF-tickling', were taken with the GEN2 trap before upgrading to the RF circuit configuration described in Appendix \ref{app_RFsetup}.
    }
    \label{fig_trapfreqvsVrf}
\end{figure}

In Fig.~\ref{fig_trapfreqvsVrf}, we observe the linear dependence of the radial trap frequencies versus the voltage pickoff from the resonator, which is proportional to the RF voltage on the RF electrodes.
Because the RF voltage is challenging to measure independently and accurately, as mentioned in Sec.~\ref{sec_trapfreqMM}, we leave it as a fitting parameter for the radial frequency simulation. Setting the DC voltages applied to the trap as fixed parameters, we use the RF voltage as the only fitting parameter for the GEN3-250$\mu$m trap simulation (blue dashed lines in Fig.~\ref{fig_trapfreqtrends}) to compare with the measurements (blue triangles), $V^{\rm{fit}}_{\rm{RF}}=482.85\,\rm{V_{pk}}$ with  $V^{\rm{fixed}}_{\rm{twist}}=-2\,\rm{V}$.

For the GEN2 trap, like the GEN3-250$\mu$m trap, the HF and LF modes at each position were taken in succession before moving to the next position. To match the measured frequencies, we find the `twist' voltage also needs to be fit with $V^{\rm{fit}}_{\rm{RF}}=377.80\,\rm{V_{pk}}$ and  $V^{\rm{fit}}_{\rm{twist}}=-0.25\,\rm{V}$, noting a +0.05 V discrepancy from the applied `twist' voltage used in the experiment (red circles in Fig.~\ref{fig_trapfreqtrends}).

For GEN3-200$\mu$m (black squares), to account for drifts in the RF voltage of $\sim\pm$0.1\% over the course of the measurement, after each point we return the ion to a reference position and measure the secular frequency there. We measure all positions for the LF mode before any  for the HF mode. The data shown is the difference between the measured frequency at each position and the reference frequency taken just after, added to the frequency measured at the trap center. We fit only the RF voltage, but due to the drift, we fit the voltage independently for the HF ($V^{\rm{fit}}_{\rm{RF}}=316.06\,\rm{V_{pk}}$) and LF ($V^{\rm{fit}}_{\rm{RF}}=315.78\,\rm{V_{pk}}$) modes with $V^{\rm{fixed}}_{\rm{twist}}=1.28\,\rm{V}$.

\subsection{\label{app_heatingratetechnical} Technical noise in heating rates}

\begin{figure}[t!]
    \centering
    \includegraphics[width=\columnwidth]{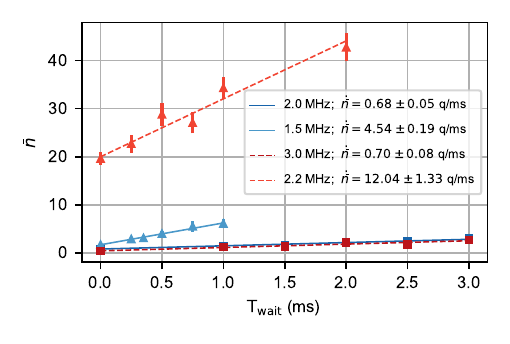}
    \caption{ 
    Motional heating rates of the radial modes measured in the GEN2 trap. Red (blue) measurements were taken at $\sim 480\, \rm{V_{pk}}$ ($\sim 377 \, \rm{V_{pk}}$) RF voltages and $\, V_{\rm{twist}}=-6\,\rm{V}$ ($\, V_{\rm{twist}}=-2.5\,\rm{V}$). Triangles and squares represent the LF and HF radial
    modes, respectively. All measurements involved extracting $\bar n$ from a thermal fit to the motional sideband evolution as a function of wait time, except for the plot with red triangles which used a thermal fit to the carrier evolution.
    }
    \label{fig_GEN2heatingrates}
\end{figure}

The GEN2 trap exhibited a much higher single-ion heating rate than the GEN3-250$\mu$m trap (see Fig. \ref{fig_GEN2heatingrates}). These heating rates were measured by probing the evolution of either the motional sideband or the carrier transition.

In Sec. \ref{sec_heatingrate_trends}, we report an anisotropy in the heating rates of the radial modes of the GEN3-250$\mu$m trap, depending on whether the mode projects more towards the DC or RF blades; we referred to these modes as the 'hot' and 'cold' modes, respectively. In addition to the peak in the single-ion heating rate spectrum of the `hot' mode, the heating rate fluctuates as well. For example, occasionally it fluctuates between 10 q/s to 50 q/s ($\omega/2\pi\sim3$ MHz, $V_{\text{twist}}=-2$ V), another indication of a technical noise source. In contrast, the heating rate of the `cold' mode fluctuates over $\sim$1-2.5 q/s from weeks to months. The measurements described in Sec.~\ref{sec_heatingrate_trends} were taken over a span of months where the heating rates were consistently the lowest. However, in Sec.~\ref{sec_coherentmanip}, the motional Ramsey coherence measurements were taken with increased technical noise on the `hot' mode. To investigate the source of the technical noise that dominates the spectral noise profile in the `hot' mode, we implement the following.

\begin{figure}[t!]
    \centering
    \includegraphics[width=\columnwidth]{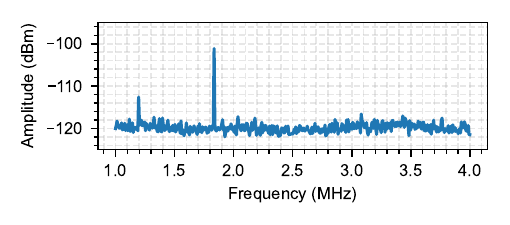}
    \caption{ 
    Power spectral density measured, using a 10:1 probe and a battery-powered spectrum analyzer, at the output of the filter box going to the electrodes, after detaching the input DAC sources. The measurement resolution bandwidth was 100 Hz.
    }
    \label{fig_DCfilterspectrum}
\end{figure}

We detach the connection from the digital-to-analog converters (DACs) to the DC and RF electrodes from the input of the filter box to rule out noise from the DACs. However, we observe similar heating rates, as discussed in Sec. \ref{sec_heatingrate_trends}. Increasing the low-pass filter's strength also did not reduce the heating rates. In fact, using a spectrum analyzer, we measure a peak in the power spectral noise profile at $\sim\,1.83$ MHz (see Fig.~\ref{fig_DCfilterspectrum}), significantly close to the radial mode frequency where the heating rate was the highest for the `hot' mode. However, converting the power spectral noise to electric field noise leads to a heating rate of several orders of magnitude higher than that measured, indicating a probe measurement error in the magnitude of the power spectral noise.

The presence of this measured noise profile without the DACs connected could indicate its origin in the experiment's ground, thereby limiting the spectral noise floor and the minimum heating rate. Noise from the ground can affect the heating rate spectrum anisotropically because the noise has a lower impedance path to the trap through the many capacitors of the $\pi$-filters in the DC electrodes' circuit, as compared to the RF electrodes' circuit which not only has fewer capacitors but also a resonator that acts as a narrow band-pass filter with bandwidth $\sim\,$100 kHz. We leave further improvements in the motional heating rate to future work.

\subsection{\label{app_motionalcoherence} Motional coherence}

\subsubsection{\label{app_derivT2} Deriving $T_2$ from a damped quantum harmonic oscillator}
The lower bound on the decoherence rate of a two-level spin system with an excited state lifetime of $T_1$ is $\Gamma^{\rm{spin}}\equiv1/T_2=1/(2\,T_1)$ \cite{Nielsen2011}. However, the decoherence rate scales differently in a quantum harmonic oscillator (QHO) in contact with a thermal bath, such as the motional modes of a trapped-ion system experiencing heating. This is primarily because the motional heating redistributes the coherence and population terms of the density matrix, $\hat\rho$, across Fock states over time. 
To estimate the decoherence scaling, $\Gamma^{\rm{tot}}$, due to both amplitude-based ($\Gamma^{\rm{a}}$) and phase-based ($\Gamma^{\rm{p}}$) decoherence, we consider a QHO in contact with an amplitude and phase reservoir similar to Ref. \cite{Brownnutt2015,turchette2000decoherence} following the Lindblad master equation:

\begin{subequations}
    \begin{align}
       \begin{split}
       \dot{\hat\rho}=\overbrace{\dot{\hat\rho}^{\rm{a}}}^{\rm{Amplitude\, noise}}+ \overbrace{\dot{\hat\rho}^{\rm{p}}}^{\rm{Phase\, noise}},
       \end{split}
       \\
       \begin{split}
        &\dot{\hat\rho}^{\rm{a}}= \frac{\gamma_a\, \bar{n}}{2}(2\hat{a}^\dagger\hat\rho \,\hat{a}-\{\hat{a}\,\hat{a}^\dagger\,,\hat\rho\})\\
       &\quad + \frac{\gamma_a\, (\bar{n}+1)}{2}(2\hat{a}\hat\rho \,\hat{a}^\dagger-\{\hat{a}^\dagger\hat{a}\,,\hat\rho\})\,,
       \\
       \end{split}
       \\
       \begin{split}
       & \dot{\hat\rho}^{\rm{p}}=\frac{\gamma_p}{2}(2\hat{a}^\dagger\hat{a}\,\hat\rho \hat{a}^\dagger\hat{a}\,-\{(\hat{a}^\dagger\,\hat{a}\,)^2,\hat\rho\})\,.
       \\
       \end{split}
    \end{align}
    \label{eq:T2deriv_rho}
\end{subequations}

Here, $\hat a$ and $\hat a^\dagger$ are the annihilation and creation operators of the bosonic motional mode connected to a thermal reservoir of average quanta, $\bar n=1/(e^{\hbar\omega/k_b T }-1)$. The heating rate of the QHO prepared near the ground state is $\dot{\bar n}=\gamma_a\bar n$, and $\gamma_p$ is the pure dephasing rate. We ignore coherent interactions from the Hamiltonian because we are interested only in decoherence in the absence of any dressing, as in the Ramsey interrogation time.

After initializing the QHO in an equal superposition of two Fock states ($\ket{n}, \ket{m}$), and taking the inner product using $\bra{n}$ and $\ket{m}$ in Eq. (\ref{eq:T2deriv_rho}), the coherence terms $\rho_{n,m}$ of the density matrix evolve as:

\begin{subequations}
    \begin{align}
       \begin{split}
       \dot{\rho}_{n,m}=\dot{\rho}_{n,m}^{\rm{a}}+ \dot{\rho}_{n,m}^{\rm{p}}\,,
       \end{split}
       \\
       \begin{split}
        &\dot{\rho}_{n,m}^{\rm{a}}= \frac{\gamma_a\, \bar{n}}{2}(2\sqrt{n\,m}\,\rho_{n-1,m-1} - (n+m+2)\rho_{n,m})\\
       &\quad + \frac{\gamma_a\, (\bar{n}+1)}{2}(2\sqrt{(n+1)(m+1)}\rho_{n+1,m+1} \\ &\quad -(n+m)\rho_{n,m})\,,
       \end{split}
       \\
       \begin{split}
       & \dot{\rho}_{n,m}^{\rm{p}}=\frac{\gamma_p}{2}(2\,n\,m\, \rho_{n,m} - n^2 \rho_{n,m} - m^2 \rho_{n,m})\,.
       \\
       \end{split}
    \end{align}
    \label{eq:T2deriv_rho_nm}
\end{subequations}

Assuming most of the coherence remains in $\rho_{n,m}$, we can approximate Eq. (\ref{eq:T2deriv_rho_nm} b,c) to only involve terms associated with $\rho_{n,m}$. This leads to :

\begin{subequations}
    \begin{align}
       \begin{split}
        &\dot{\rho}_{n,m}^{\rm{a}}\approx -[\frac{\gamma_a\, \bar{n}}{2} (n+m+2)\\
       &\quad + \frac{\gamma_a\, (\bar{n}+1)}{2}  (n+m)]\, \rho_{n,m}\,,
       \end{split}
       \\
       \begin{split}
       & \dot{\rho}_{n,m}^{\rm{p}}\approx-\frac{\gamma_p}{2}(n-m)^2\rho_{n,m}\,.
       \\
       \end{split}
    \end{align}
    \label{eq:T2derivapprox_rho_nm}
\end{subequations}

The above equations hint an exponential scaling of $\rho_{n,m}$ with time, which aligns well with the conventional picture of decoherence. However, this approximation for the amplitude-based decoherence is valid only for low phonon excitation, in the limit of $\dot{\bar n} \,t\ll1$, which would otherwise follow Eq. (7) in Ref. \cite{turchette2000decoherence} due to significant mixing of the coherence and population terms.
Under the approximation of low phonon excitation, the total decoherence rate scales as :
\begin{equation}
    \Gamma^{\rm{tot}}\approx \underbrace{\dot{\bar n}(n+m+1)\, +\,\frac{\dot{\bar n}(n+m)}{2\,\bar{n}_{300\rm{K} }}}_{\Gamma^{\rm{a}}}
    \,+\,\underbrace{\frac{\gamma_{\rm{ph}}}{2}\,(n-m)^2}_{\Gamma^{\rm{p}}}
\end{equation} 
The pure amplitude-based decoherence rate agrees with the scaling derived in Ref. \cite{turchette2000decoherence} (Eq.(7),(38)) within the limit mentioned. Contrary to the amplitude-based decoherence, the phase-based decoherence rate scales exponentially with the square of the energy difference between the two Fock states. These results justify the use of an exponential to fit the motional Ramsey coherence measurements discussed in Sec.~\ref{sec_coherentmanip}.
For Fock states $\ket{n=0}$ and $\ket{m=1}$,
\begin{equation}
T_2\approx1/(2\dot{\bar n}+\gamma_{\rm{ph}}/2)\, .
\label{eq_motionalT2_01}
\end{equation}

\begin{figure}[t!]
    \centering
    \includegraphics[width=\columnwidth]{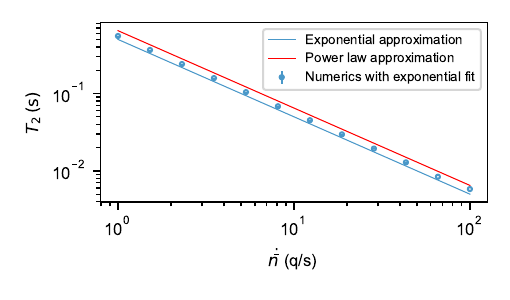}
    \caption{ 
   Scaling of $T_2$ for a given heating rate. We compare the $1/e$ time of $T_2$ from the exponential scaling approximation of the coherence time in Eq.~\eqref{eq_motionalT2_01} (blue line), and the power law approximation in Eq.(38) (red line) of Ref.~\cite{turchette2000decoherence} to exact numerics (blue dots). The numerics for $N_{\rm{phonon}}^{\rm{cutoff}}=30$ levels, are fit using an exponential function up to the motional Ramsey wait time where the contrast is 0.5 to extract $T_2$ ( maximum wait time= 400 ms). Explicit motional dephasing is absent here.
   }
    \label{fig_T2_scaling}
\end{figure}

We numerically verify the scaling of the amplitude-based dephasing with the analytically derived scaling involving the heating rate, without pure dephasing, in Fig.~\ref{fig_T2_scaling}, using QuTiP \cite{JOHANSSON2013}. The $T_2$ time of the numerical simulations (blue dots) are upper bounded by Eq. (38) of Ref.\cite{turchette2000decoherence} (red line) and lower bounded by Eq.~\eqref{eq_motionalT2_01} of this text (blue line).

When $\dot{\bar n}\gg\gamma_{\rm{ph}}$, as in the case of the LF mode, $\dot{\bar n}\approx50\,\rm{q/s}$ (discussed in Sec.~\ref{sec_coherentmanip}), we find good agreement of the motional Ramsey decoherence measurement with the scaling of Eq.~\eqref{eq_motionalT2_01} which shows $T_2\approx10 \,\rm{ms}$. The HF mode, $\dot{\bar n}\approx1.6\,\rm{q/s}$, on the other hand is dominated by motional dephasing,  $\gamma_{\rm{ph}}\approx2\pi\cdot 2.3\, \rm{Hz}$, which was fitted by using a numerical simulation. Using this fitted dephasing rate in Eq.~\eqref{eq_motionalT2_01} results in $T_2=95.9\,\rm{ms}$, which is in agreement with the experimentally observed value. 

\subsubsection{\label{app_motionalphasescans}Motional Ramsey phase scans with frequency drifts}

\begin{figure}[t!]
    \centering
    \includegraphics[width=\columnwidth]{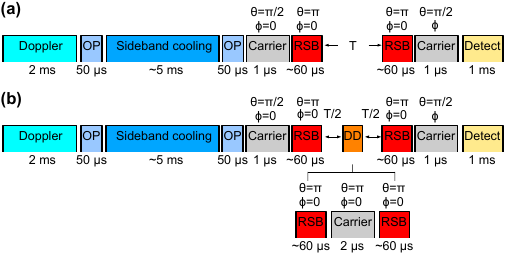}
    \caption{ 
  Pulse sequence of motional Ramsey experiments {\bf (a)} without and {\bf (b)} with a motional echo pulse (referred to as `DD' in the figure, with Ramsey interrogation time, $T$. Here `OP' is the optical pumping pulse; `Carrier' and `RSB' are the coherent Raman pulses driven on the ground-state clock qubit and the red-motional sideband, respectively. $\theta$ is the unitary rotation angle, and $\phi$ is the Raman beatnote's phase. All coherent operations are implemented using a stimulated Raman process.
   }
    \label{fig_motionalT2pulseseq}
\end{figure}

\begin{figure}[t!]
    \centering
    \includegraphics[width=\columnwidth]{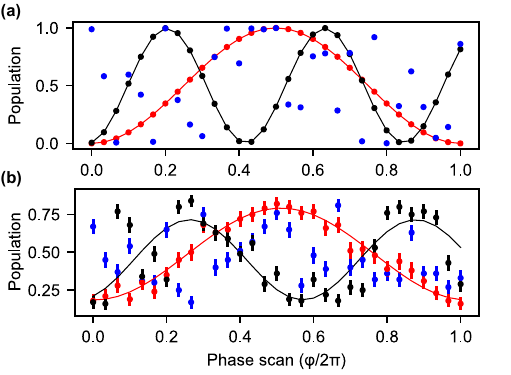}
    \caption{ 
    {\bf (a)} Simulation of the motional Ramsey phase scan in the presence of a 0.1 Hz/s drift in the radial mode, over the duty cycle of the experiment with $T_{\rm{wait}}=$ 20 ms. The red (black) plot assumes without (with) frequency drift in the linear scan of Ramsey phase. The blue plot assumes frequency drift in the randomized scan of the Ramsey phase.
    {\bf (b)} Motional Ramsey phase scan measurements of the HF mode as in Fig.~\ref{fig_coherentdynamics1} b,c, with (red) and without (black) a motional echo in the presence of frequency drift for $T_{\rm{wait}}=$ 20 ms.
    The blue measurements were taken with a randomized Ramsey phase at $T_{\rm{wait}}=$ 25 ms.
   }
    \label{fig_phasescans-2Vtwist}
\end{figure}

\begin{figure}[t!]
    \centering
    \includegraphics[width=\columnwidth]{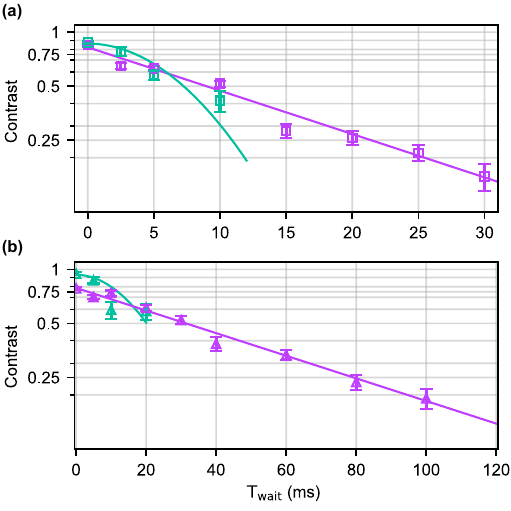}
    \caption{ 
    Motional Ramsey coherence time of the {\bf (a)} HF and {\bf (b)} LF modes in the $V_{\text{twist}}=+2\,\text{V}$ configuration contrary to $V_{\text{twist}}=-2\,\text{V}$ in Fig. \ref{fig_coherentdynamics1} b,c. The turquoise and purple plots are without and with a motional echo, respectively. Measurements without the motional echo are included in the fit up to the point where the period of the Ramsey phase scan significantly deviates from $2\pi$. The spin Ramsey coherence time of the experiment using the Raman transition is $\approx500\,\rm{ms}$
   }
    \label{fig_motionalcoherence+2Vtwist}
\end{figure}

In Fig.~\ref{fig_phasescans-2Vtwist}, we compare the observed and simulated motional Ramsey phase scans of the HF mode, with and without a motional echo, to depict the deviation of the phase scan from the typical $2\pi$ period because of the motional frequency drift over the experiment's duty cycle (see Fig.~\ref{fig_motionalT2pulseseq}). 

Following the same motional Ramsey protocols, the motional coherence characteristics are exchanged between the two motional modes upon inverting the `twist' voltage (see Fig.~\ref{fig_motionalcoherence+2Vtwist}). For these measurements, $\dot{\bar n}\sim2.5\, \rm{q/s}$ for the LF mode ({\bf (b)} triangles)
, and $\dot{\bar n}\sim25\, \rm{q/s}$ for the HF mode ({\bf (a)} squares). We fit the motional coherence measurements of both the modes without (with) a motional echo to a Gaussian (exponential) function. We measure $T_2^*=29(3)$ ms for the LF mode, and $T_2^*=11(2)$ ms for the HF mode. With the motional echo, we measure $T_2=69(4)$ ms for the LF mode, and $T_2=18(2)$ ms for the HF mode. These results indicate that the anisotropic heating rates, along with motional dephasing, are correlated with the anisotropic motional coherence of the radial modes as well.

\section{\label{app_trapsim} Trap simulations}

During the trap design process, we modeled the potential and charge distribution on the electrodes using the finite element analysis (FEA) software, COMSOL. The charge distribution of GEN1 was concentrated near the RF-DC trenches, hence we changed the trench geometry in GEN2 to lower the trench capacitance and thereby reduce the local power dissipation at the trenches. After measuring the excess micromotion in GEN2 (see Fig.~\ref{fig_MMtrends}), we found that the shape of the micromotion could be largely predicted by the FEA results, if the entire trap was accounted instead of just the blades. The simulation showed that the endcaps on one side of the trap had $\sim\,$0.01 pF more capacitance to the RF blades than the endcaps on the other side, leading to asymmetric RF voltage on the endcaps. This led to the more symmetric design of GEN3, where the capacitance asymmetry is 10x lower, with the remaining asymmetry most likely due to residual asymmetries from the greater proximity of the RF wires to the DC wires on one side of the trap compared to the other.

We also use the FEA result to find the trap depth. For the DC axial confinement, the trap depth is given by the charge multiplied by the potential difference from the DC potential minimum to maximum. Under normal operation the minimum is at the the trap center, where the endcaps raise the potential by \mbox{$0.07 \: (0.04) \,\mathrm{V}/\mathrm{V}_{\mathrm{endcap}}$} for GEN3-250$\mathrm{\mu}$m (GEN3-200$\mathrm{\mu}$m). The potential maximum occurs near the center of the endcaps at $0.5\, \mathrm{V}/\mathrm{V}_{\mathrm{endcap}}$, where the factor of $0.5$ comes from the potential on the axis being the average of the endcaps voltage and  0 V$_{\mathrm{DC}}$ on RF blades. Hence, the trap depth is $0.43 \: (0.46) \,\mathrm{eV}/\mathrm{V}_{\mathrm{encdap}}$ for  GEN3-250$\mathrm{\mu}$m (GEN3-200$\mathrm{\mu}$m).

\begin{figure}[t!]
    \centering
    \includegraphics[width=\columnwidth]{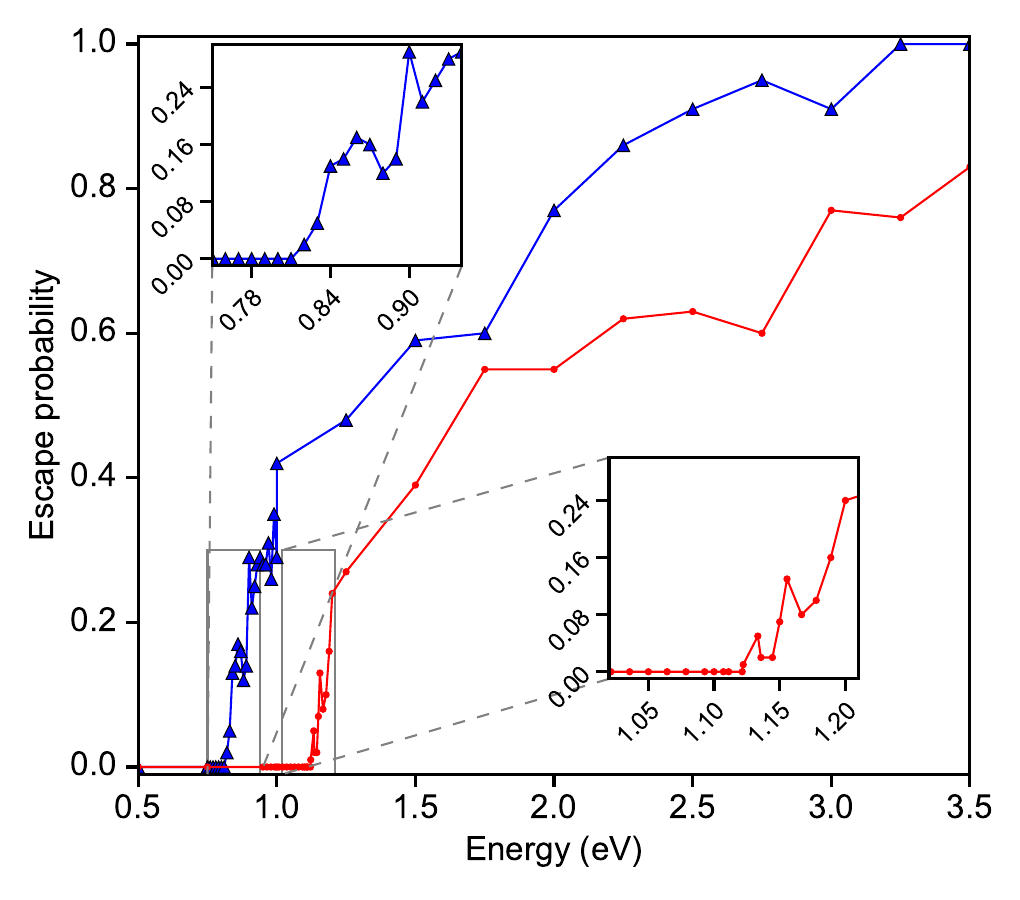}
    \caption{Monte-Carlo simulation of the escape probability within 50 secular frequency cycles versus $^{171}$Yb$^+ $ energy at $\omega_{\rm{}}/2\pi=3$ \text{MHz}, $\Omega_{\rm{RF}}/2\pi$ = 23 MHz, 100 shots per energy with random initial velocity direction. The blue curve is for GEN3-200$\mu\rm{m}$ and the red curve is for GEN3-250$\mu\rm{m}$.
    }
    \label{trap_depth_figure}
\end{figure}

 For the radial confinement, rather than using the pseudopotential, we simulate the motion of an ion to account for the anharmonicity near the blade tips. We perform a Monte-Carlo simulation where the ion begins in the trap center with a random velocity and is allowed to evolve for secular motion periods. If the ion begins with kinetic energy lower than the trap depth, it never escapes past the trap.  We set the RF voltage to 3 MHz secular frequency. No DC voltages are simulated, and only the radial plane is considered.  Fig.~\ref {trap_depth_figure} shows the probability of escape within 50 secular frequency cycles as a function of initial kinetic energy. We expect that if the ion is allowed to evolve for longer times, the curves would asymptote to step functions with the rising step at the trap depth. Due to the finite number of cycles, we bound the trap depth, $D$, to $0.81\,\text{eV} <D<0.82\,\text{eV}$ for GEN3-200$\mu\rm{m}$ and  $1.21\,\text{eV}<D<1.23\,\text{eV}$ for GEN3-250$\mu\rm{m}$. The larger trap depth for GEN3-250$\mu\rm{m}$ is due to the higher voltage necessary to maintain the same secular frequency.

\section{\label{app_RFsetup} RF circuit}

The RF voltage from a low-phase noise signal generator (Rohde \& Schwarz SMB100B) is amplified using an air-cooled RF amplifier (ZHL-20W-13+) and a home-built helical resonator \cite{Siverns2012OnResonators, deng2014modified} to provide an impedance-matched, high-RF voltage at the trap. The resonator has two secondary RF coils, one for each RF blade, which are capacitively shorted to have the same RF phase, while still allowing independent DC offsets through bias-tees. The quality factor, Q, of the RF circuit is typically between 150 and 230.
The RF voltage on the trap is sampled from a $\approx 2\%$ voltage pickoff of an inductive loop (3/4 turn, 50 mm diameter, bare copper, 14 AWG) inside the resonator, which immediately follows a bridge rectifier, with resistor-based temperature matching \cite{Johnson2016}, outside the resonator. This provides about $\approx$ 25\% conversion efficiency ($\rm{V_{DC}/V_{pk-pk}}$) for $50\,\Omega$ impedance matching at the rectifier input, and $\approx 50\%$ without input impedance matching (operating condition). The rectifier's efficiency has a measured temperature sensitivity of $\sim$ -500 ppm/$^{\circ}$C.
The pickoff voltage is stabilized using a Proportional-Integral, Newport lockbox (LB1005-S) with respect to an 18-bit DAC voltage source (EVAL-AD5781), by providing feedback through a voltage variable attenuator (ZX73-2500+, non-powered) at the amplifier input. 
The two DC-biases for each RF electrode pass through a `bias-tee', consisting of a $1\,\mu$F capacitor to ground and the resonator's secondary coil ($\sim1.7\,\mu$H) in series, before entering the trap. The RF electrode biases have two $\pi-$filters in series, compared to the DC electrode biases that have only one $\pi-$filter per electrode.

All connections in the circuit use stiff and short (mostly), SMA cables (Minicircuits 141-xMSM+) to minimize the effect of residual ground loops and minimize sensitivity to mechanically-induced voltage drifts in the circuit.
We minimize the number of ground loops in the RF circuit by passing the AC-mains line power through a 1:1 transformer, and then connecting all the chamber-associated circuitry (RF, DC, ion pumps, etc.) on a power strip connected to this transformer. The ground is isolated for each device connected to this strip, forming a `star' configuration with respect to the vacuum chamber. The chamber is then connected to a ground plane via a copper strip, serving as the single ground connection. For the measurements reported here, the system was not in a complete `star' configuration due to a residual short to ground through the chamber's external support. This may have contributed to the observed technical in heating rates.

\section{\label{sec_technicalMS} M\o lmer-S\o rensen Interaction }

\begin{figure}[t!]
    \centering
    \includegraphics[width=\columnwidth]{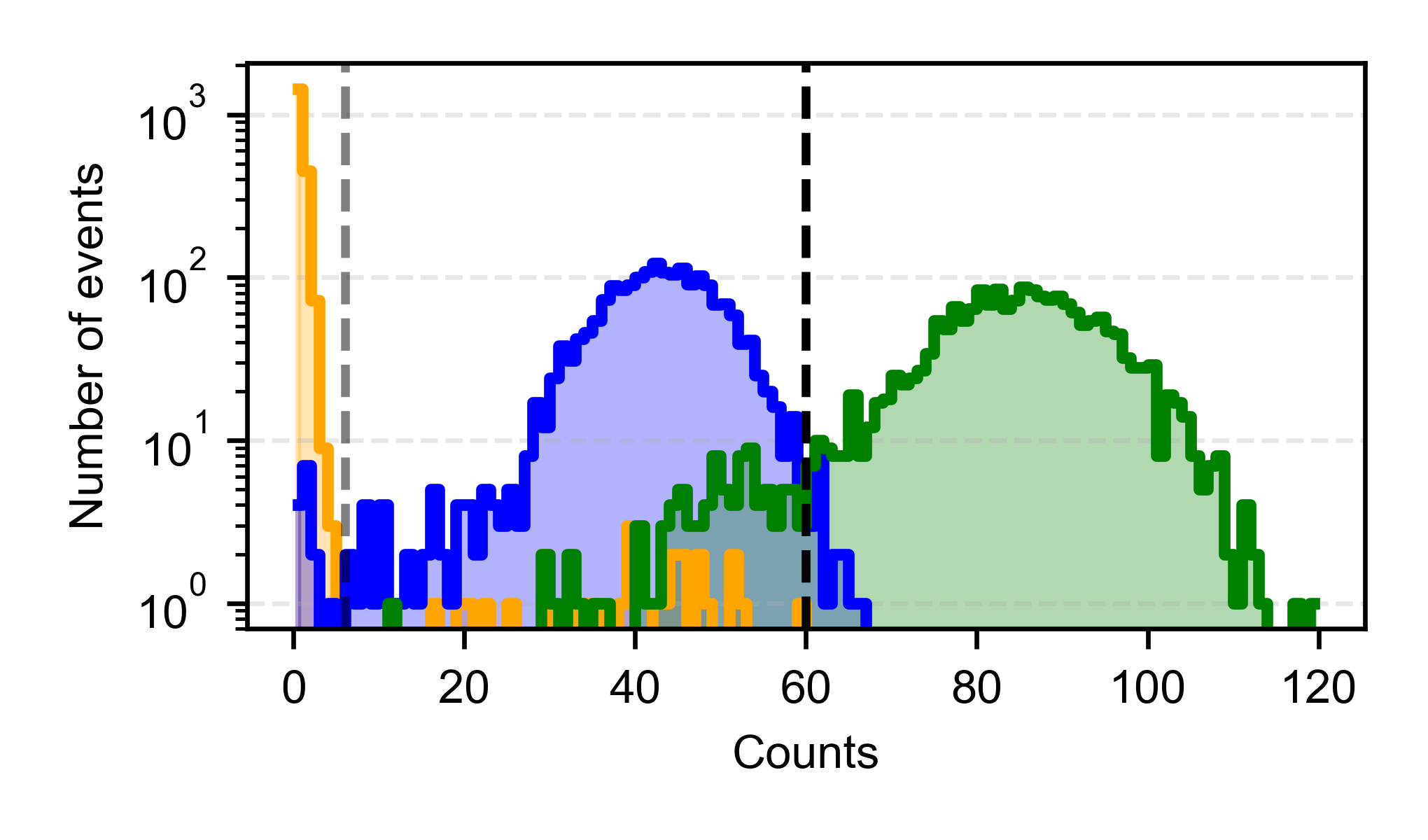}
    \caption{ 
    Histogram of fluorescence collected after preparing two \Ybf ions in: $\ket{\downarrow_{{z}}\downarrow_{{z}}}$ (orange), $\ket{\downarrow_{{z}}\uparrow_{{z}}}$ or $\ket{\uparrow_{{z}}\downarrow_{{z}}}$ (blue), and $\ket{\uparrow_{{z}}\uparrow_{{z}}}$ (green), for 2000 shots. The two threshold values are 6 (gray dashed line) and 60 (black dashed line) counts.
    }
    \label{fig_MS_histogram}
\end{figure}

Before implementing the M\o lmer-S\o rensen interaction on two ions, the SPAM fidelity is measured using a PMT for 2000 repetitions, resulting in the fidelity matrix:

\begin{equation}
M_{\text{SPAM}} = 
\begin{pmatrix}
0.9865(26) & 0.0070(19) & 0 \\
0.0135(26) & 0.9840(28) & 0.0475 (48) \\
0 & 0.009(21) & 0.9525 (48)
\end{pmatrix}
\label{mat:SPAM}
\end{equation}

obtained from the histograms in Fig.~\ref{fig_MS_histogram}. Error bars denote the 68.3\% credible interval derived from 10,000 Dirichlet Monte Carlo samples from the distribution in Fig.~\ref{fig_MS_histogram}. The thresholds are optimized for maximizing the product of the fidelities to prepare each of the three distinguishable states (ordered by row from top to bottom in Eq.~\eqref{mat:SPAM}): 
$\ket{\downarrow_{{z}}\downarrow_{{z}}}$, $\ket{\downarrow_{{z}}\uparrow_{{z}}}$ or $\ket{\uparrow_{{z}}\downarrow_{{z}}}$, and $\ket{\uparrow_{{z}}\uparrow_{{z}}}$.
The maximum fidelity for the two qubit XX gate was obtained by parking at $t_{\rm{gate}}=2\pi/\delta_{gate}$ and scanning the amplitude of the bichromatic beatnote, where the optimal gate amplitude is achieved when the population reaches 0.5 for both $\ket{\downarrow_{{z}}\downarrow_{{z}}}$ and $\ket{\uparrow_{{z}}\uparrow_{{z}}}$. The parity is scanned after preparing $\ket{\Phi}$ with these parameters. Since the Bell state preparation fidelity is lower bounded by the contrast of the parity scan \cite{Nielsen2011}, we predict the Bell state preparation fidelity to be $\geq 99.3^{+0.7}_{-1.5} \%$ , after SPAM correction.

The MLE \cite{eliason1993MLE} fit consists of the following: \textbf{\emph{i)}} Due to the conservation of total population and the definition of parity \cite{sackett2000_4ions}, we consider the net even parity population, $P_{\rm{even}}=P_{\ket{\uparrow_{{z}}\uparrow_{{z}}}}+P_{\ket{\downarrow_{{z}}\downarrow_{{z}}}}$, as the random variable and extract the Beta distribution bounded errors (68.3\% confidence interval) from its thresholded counts. \textbf{\emph{ii)}} The parity fit is a sinusoidal function from which the fit's populations $p_{\ket{\uparrow_{{z}}\uparrow_{{z}}}}$, $p_{\ket{\downarrow_{{z}}\downarrow_{{z}}}}$, and $p_{\ket{\downarrow_{{z}}\uparrow_{{z}}} \rm{or}\ket{\uparrow_{{z}}\downarrow_{{z}}}}$ are extracted. 
The populations are propagated by $M_{\rm{SPAM}}$ only for SPAM correction, resulting in $\vec{p}\,'=M_{\rm{SPAM}}\cdot \vec{p}$, instead of $\vec{p}\,'= \vec{p}$ without the SPAM correction.
\textbf{\emph{iii)}} 
The cost function, 
\begin{equation}
C=\sum_{k}^K\sum_{\substack{\psi\in\{\uparrow_{{z}},\downarrow_{{z}}\}^{\otimes2}}}-N_{\psi,k}\,\text{log}(p\,'_{\psi,k}),
\end{equation} 
is minimized, where $p\,'_{\psi,k}$ is the fit's population per distinguishable state for each scan point $k$ of the total $K$ scan points, and $N_{\psi,k}$ is the number of times the state, $\ket{\psi}$, was detected based on the thresholds provided.
\textbf{\emph{iv)}} The fit's amplitude determines the parity contrast. The fitting parameter errors are obtained from computing the inverse of the loss function's Hessian matrix, which assumes a Gaussian error distribution. However, because the parity contrast cannot exceed 1, this parameter bound causes the error bars to follow an asymmetric Gaussian distribution for the $68.3\%$ confidence interval.

This technique is more accurate for extracting SPAM-corrected metrics from limited statistics, as it does not fit the SPAM-corrected parity measurement itself; instead, it accounts for the correction in the fit and relies on the measurement's distribution rather than assuming one. For this reason, the purple points in Fig.~\ref{fig_MSevol} are the SPAM corrected parity measurements, using $M^{-1}_{\rm{SPAM}}$, that are not used in the fit, but are shown only for comparison.

\begin{table*}[h!]
    \centering
    \caption{Legend of references of the single-ion heating rates in Fig. \ref{fig_heatingrate_comparison}.}
    \label{tab_heatingrate_legend}
    \small 
    \begin{tabular}{|c|l|c||c|l|c|}
        \hline
        \textbf{No.} & \textbf{Label} & \textbf{Reference} & \textbf{No.} & \textbf{Label} & \textbf{Reference} \\
        \hline
        1 & Sandia; HOA 2.1.1 (Cetina, 2022) & \cite{Cetina2022} & 33 & MIT; surface (Wang, 2010) & \cite{wang2010demonstration} \\
        2 & Sandia; Enchilada (Sterk, 2024) & \cite{sterk2024multi} & 34 & MIT; surface (Stuart, 2019) & \cite{stuart2019chip} \\
        3 & Sandia; Phoenix (Sun, 2024) & \cite{sun2024quantum} & 35 & MIT; surface (Labaziewicz, 2019) & \cite{labaziewicz2008temperature} \\
        4 & PTB; 3D chip (Jordan, 2025) & \cite{jordan2025scalable} & 36 & Oxford; rod (Lucas, 2007) & \cite{lucas2007longlivedmemoryqubitlowdecoherence} \\
        5 & Innsbruck; rod (Rohde, 2001) & \cite{rohde2001sympathetic} & 37 & Oxford; surface (Allcock, 2010) & \cite{allcock2010implementation} \\
        6 & Innsbruck; blade (Benhelm, 2007) & \cite{benhelm2007measurement} & 38 & Oxford; surface (Allcock, 2011) & \cite{allcock2011reduction} \\
        7 & Innsbruck; 3D chip (Harlander, 2012) & \cite{harlander2012architecture} & 39 & Oxford; surface (Allcock, 2013) & \cite{allcock2013microfabricated} \\
        8 & Innsbruck; blade (Lechner, 2016) & \cite{Lechner2016EITStrings} & 40 & Oxford; 3D (Ballance, 2016) & \cite{Ballance16} \\
        9 & Innsbruck; 3D chip (Keisenhofer, 2023) & \cite{Kiesenhofer2023} & 41 & Oxford; surface (Weber, 2024) & \cite{weber2024cryogenic} \\
        10 & UCB; surface (Daniilidis, 2012) & \cite{daniilidis2011fabrication} & 42 & FOCUS; 3D (Deslauriers, 2004) & \cite{deslauriers2004zero} \\
        11 & UCB; surface (Daniilidis, 2014) & \cite{daniilidis2014surface} & 43 & FOCUS; 3D (Stick, 2006) & \cite{stick2006ion} \\
        12 & UCB; surface (Noel, 2019) & \cite{noel2019electric} & 44 & FOCUS; needle (Deslauriers, 2006) & \cite{Deslauriers2006} \\
        13 & UCB; 3D-printed (Xu, 2025) & \cite{Xu20253D} & 45 & Ulm; 3D chip (Poschinger, 2009) & \cite{poschinger2009coherent} \\
        14 & NIST; rod (Myatt, 2000) & \cite{myatt2000decoherence} & 46 & GTech; surface (Doret, 2012) & \cite{Doret2012} \\
        15 & NIST; linear (Turchette, 2000) & \cite{turchette2000heating} & 47 & GTech; surface (Guise, 2015) & \cite{guise2015ball} \\
        16 & NIST; linear (Turchette, 2000) & \cite{turchette2000heating} & 48 & Siegen; surface (Boldin, 2018) & \cite{boldin2018measuring} \\
        17 & NIST; 3D (`dual') (Rowe, 2002) & \cite{rowe2002transportquantumstatesseparation} & 49 & Sussex; blade (Weidt, 2015) & \cite{weidt2015ground} \\
        18 & NIST; 3D (`trap C') (Rowe, 2002) & \cite{rowe2002transportquantumstatesseparation} & 50 & Aarhus; rod (Poulsen, 2012) & \cite{poulsen2012efficient} \\
        19 & NIST; surface (Epstein, 2007) & \cite{epstein2007simplified} & 51 & ETH; 3D chip (Kienzler, 2015) & \cite{Kienzler2015} \\
        20 & NIST; surface (Leibrandt, 2009) & \cite{leibrandt2009demonstration} & 52 & ETH; surface (Mehta, 2020) & \cite{Mehta2020} \\
        21 & NIST; surface (Britton, 2009) & \cite{britton2009scalable} & 53 & Sydney; blade (Matsos, 2025) & \cite{matsos2025universal} \\
        22 & NIST; surface (Amini, 2010) & \cite{amini2010toward} & 54 & NPL, 3D chip (Wilpers, 2012) & \cite{wilpers2012monolithic} \\
        23 & NIST; surface (Ospelkaus, 2011) & \cite{ospelkaus2011microwave} & 55 & UMD; blade (Pagano, 2018) & \cite{Pagano2018} \\
        24 & NIST; 3D chip (Blakestad, 2011) & \cite{blakestad2011near} & 56 & UMD; rod (Carter, 2024) & \cite{carter2024ion} \\
        25 & NIST; surface (Warring, 2013) & \cite{warring2013techniques} & 57 & Tsingua; 3D chip (Wang, 2020) & \cite{Wang2020coherently} \\
        26 & NIST; surface (McCormick, 2019) & \cite{mccormick2019quantum} & 58 & Infineon; 3D chip (Auchter, 2022) & \cite{auchter2022industrially} \\
        27 & NIST; surface (Lysne, 2024) & \cite{lysne2024individual} & 59 & SNU; blade (Jeon, 2025) & \cite{Jeon2025twomodesingleshot} \\
        28 & Lincoln; surface (McConnel, 2015) & \cite{mcconnell2015reduction} & 60 & Weizmann; rod (Akerman, 2012) & \cite{akerman2012quantum} \\
        29 & Lincoln; surface (Sedlacek, 2018) & \cite{sedlacek2018evidence} & 61 & Rice; blade (So, 2024) & \cite{so2024trappedion} \\
        30 & Lincoln; surface (Niffenegger, 2020) & \cite{Niffenegger2020} & 62 & Rice; monolithic 3D (2024) & \parbox{2cm}{ This work,\\ GEN2\\{}} \\[5pt]
        31 & MIT; surface (Leibrandt, 2009) & \cite{leibrandt2009demonstration} & 63 & Rice; monolithic 3D (2025) &
        \parbox{2cm}{This work,\\ GEN3-250$\mu$m } \\
        32 & MIT; surface (Wang, 2010) & \cite{wang2010demonstration} & & & \\
        \hline
    \end{tabular}
\end{table*}

\clearpage

\bibliographystyle{Monolithic_sty}
\bibliography{references}

\end{document}